%% file: main.tex
\documentclass[aps,prev,preprintnumbers,floatfix,nofootinbib]{revtex4-1}
\input{Diagrams}

\usepackage{graphicx}
\usepackage{amsmath, amsthm, amssymb}
\usepackage{amsfonts}
\usepackage{subfig}
\usepackage{xspace}
\usepackage[dvipsnames]{xcolor}
\usepackage{paralist}
\usepackage{multirow}
\usepackage{paralist}
\usepackage{graphicx}
\usepackage{bm}
\usepackage{times}
\usepackage{slashed}
\usepackage{color}
\usepackage{epsfig}\usepackage{dcolumn}
\usepackage[percent]{overpic}
\usepackage{color}
\def\br{\begin{eqnarray}}
\def\er{\end{eqnarray}}
\def\be{\begin{equation}}
\def\ee{\end{equation}}

\begin{document}

\title{Probing the existence of a new charged vector boson decaying into heavy neutral leptons using ultra-peripheral heavy ion collisions at ATLAS}

\author{Y.M. Oviedo-Torres$^{1,2}$}
\author{Sebastian Tapia$^{3,4}$}
\author{J. Zamora-Saa$^{1,2}$}

\email{mauricio.nitti@gmail.com}

\affiliation{$^1$Millennium Institute for Subatomic Physics at High-Energy Frontier (SAPHIR), Fernandez Concha 700, Santiago, Chile. \\
$^2$Center for Theoretical and Experimental Particle Physics - CTEPP, Facultad de Ciencias Exactas, Universidad Andres Bello, Fernandez Concha 700, Santiago, Chile\\
$^3$Departamento de Física, Universidad Técnica Federico Santa María, Avenida España 1680, Edificio E, Valparaíso Chile.\\
$^4$Centro Cientifico Tecnólogico de Valparaíso (CCTVal), Universidad Técnica Federico Santa María, Valparaíso, Chile.
}

\begin{abstract}

In this paper we explore the potential of Ultra-peripheral Collisions at the LHC to investigate new physics, focusing on the production of new charged vector bosons ${V}^{\pm}$ that decay into heavy neutral leptons $N_{L}$ in the context of the Vector Scotogenic Model. We show that the ATLAS experiment, searching for dilepton+met final states via UPCs of lead ions, can directly probe the existence of new charged vector bosons in the 5 GeV $<M_{V^{\pm}}, M_{N_L}<$ 105 GeV mass range. Our analysis identifies regions in the parameter space where the signal can be distinguished from the background with high statistical significance. Within this mass range, ATLAS can exclude at 95\% C.L. specific (${M}_{{V}^{\pm}}$, ${M}_{{N}_{L}}$) mass scenarios such as (30 GeV, 20 GeV), (30 GeV, 10 GeV) and (20 GeV, 10 GeV). In a discovery scenario, ATLAS could reach a significance of 5$\sigma$ for (20 GeV, 10 GeV). In addition, we show that HL-LHC with proton-proton UPCs could explore higher mass ranges, specifically 100 GeV $< M_{V^{\pm}}, M_{N_L}<$ 350 GeV. We find that the HL-LHC can exclude this mass range with 95\% C.L., covering a larger parameter space than previous SUSY searches, and most scenarios can achieve a discovery significance of 5$\sigma$.

\end{abstract} 
    
\maketitle

\section{Introduction}
\label{introduction}

There is no doubt that the Standard Model (SM) of particle physics is the most appropriate theoretical framework to describe the fundamental interactions between elementary particles. However, there is a general consensus in the scientific community that the SM is not the final theory on fundamental interactions. It fails to address several fundamental questions, such as the nature of dark matter \cite{Bertone:2004pz, Boucenna:2012rc, Mambrini:2015sia,  Hambye:2010zb}, the asymmetry between matter and antimatter \cite{Affleck:1984fy, Sakharov:1967dj, Yoshimura:1978ex, Dimopoulos:1978kv,Ellis:1978xg}, the neutrino mass problem \cite{PhysRevLett.44.912,PhysRevD.34.1642,Ma:1998dx,Ding:2023htn}, among others \cite{Bilson-Thompson:2006xhz, Chamseddine:1982jx, Lee:2019zbu}. These open problems in fundamental physics provide an ideal scenario for developing more general theoretical frameworks or conducting more sophisticated experiments. Among the various experiments, hadron colliders play a crucial role in exploring these open problems, as they provide the required energy scale where these new interactions could occur. These machines are specifically designed to probe subatomic interactions through the direct collision of hadrons, offering the perfect environment to examine not only the predictions of the SM but also to search for signals of new physics Beyond the Standard Model (BSM). The Large Hadron Collider (LHC), for example, stands as the most powerful particle collider ever constructed for this purpose \cite{Halkiadakis:2014qda, ATLAS:2008xda, CMS:2008xjf}. With the ability to collide protons at energies of up to 13.6 TeV, the LHC has enabled experiments that have confirmed many predictions of the SM, such as the existence of the Higgs boson, which was detected in 2012 \cite{ATLAS:2012yve, CMS:2012qbp}. As we can see, the acceleration of proton beams has been crucial for exploring particle physics. However, the use of protons as collider beams presents a problem inherent to the Quantum Chromodynamics nature of hadrons, which counteracts the advantages of their use\footnote{In this work we refer to this type of collisions as conventional proton-proton collisions.}.  \newline

We know that protons are particles composed by quarks and gluons that interact through the strong nuclear force. This necessarily implies that conventional proton-proton collisions generate a significant amount of irreducible hadronic background, making it difficult the identification of many interesting processes that could reveal signals of new physics, especially those that occur at energies below the available partonic center-of-mass energy. For this reason, it is essential to mitigate the impact of the background contamination on relevant signals using advanced data detection and analysis techniques, as well as improvements in reconstruction and event selection algorithms, in order to properly separate the signals of interest from the background. A strategy that has been little explored in the literature which can help minimize this hadronic background is the search for new physics through Ultra-Peripheral Collisions (UPCs). The main characteristic of this type of collision is that the beam components (such as lead ions in the LHC experiment, for example \cite{CMS:2011vma, ATLAS:2017fur, ATLAS:2020epq, ATLAS:2022kas,ATLAS:2025aav,ATLAS:2025nac,ATLAS:2025rcd,Ciesla:2024pvn,ATLAS:2024mvt,ATLAS:2024nzp}) do not fragment during the collisions, but pass very close to each other, interacting through their electromagnetic fields. These fields can be very intense in heavy ions due to the high electric charge of the nuclei (for example, the lead nucleus has an electric charge of +82e, in contrast to the proton, whose electric charge is only +1e). These electromagnetic interactions allows for the generation of interesting processes without hadronic contamination, in which new particles can be produced, such as charged vector bosons and heavy neutral leptons (HNLs), which are studied in this work.   \newline

Motivated by the reduced background produced in UPCs compared to conventional proton-proton collisions, this paper studies the feasibility of producing new charged vector bosons that decay into HNLs, predicted in the Vector Scotogenic Model (VSM), for the current stage of the ATLAS Experiment using Pb-Pb collisions (PbPb-UPC) and the future high-luminosity configuration of the LHC (HL-LHC) using proton collisions (pp-UPC). For that purpose, this paper is organized as follows: In Section \ref{introduction}, we present the introduction to this work; in Section \ref{upc}, we discuss general aspects of ultra-peripheral collisions; in Section \ref{vsmmodel} we describe the main features of the Vector Scotogenic Model; In Section \ref{upc_atlas} we discuss the possibility of observing signals of new physics at the current stage reached by the ATLAS experiment with UPCs of lead ions. In Section \ref{upc_14LHC_pp} we extend our analysis to the HL-LHC using UPCs of protons. Finally, in Section \ref{conclusions}, we summarize our conclusions.  \newline

\section{Probing new physics with Ultra-peripheral Collisions}
\label{upc}

It is well-established that conventional proton-proton collisions, such as those taking place most of the time at the LHC, represent one of the primary avenues for testing phenomena predicted by the SM and for directly or indirectly constraining many BSM theories. This is largely due to the nature of the proton itself. As a hadron, most of the interactions in these collisions are dominated by the strong interaction between its partonic components. These partonic interactions give rise to a rich phenomenology that can be used to constrain, for example, the properties of hypothetical particles. However, these types of collisions also present a significant challenge: the enormous amount of hadronic background generated, which can obscure many signals of new physics.  \newline

An alternative and relatively underexplored approach that significantly reduces the QCD background and creates an ideal environment for searching for new particles is the use of Ultra-peripheral Collisions. In UPCs, the electromagnetic field of one nucleus induces inelastic transitions in the other, resulting in processes that share the same transition matrix elements as those induced by quasi-real photons \cite{Bertulani:2005ru, Bertulani:2023tgh}. This implies that the majority of interactions in UPCs are governed by electromagnetic interactions rather than strong interactions. Furthermore, in most UPC processes, there is no fragmentation of the beam constituents. These two characteristics lead to cleaner final states, which can result in improved signal-to-background ratios for new physics searches compared to conventional proton-proton collisions.  \newline 

\begin{figure}[ht]
    \begin{minipage}[b]{0.6\textwidth}
        \begin{overpic}[width=\textwidth]{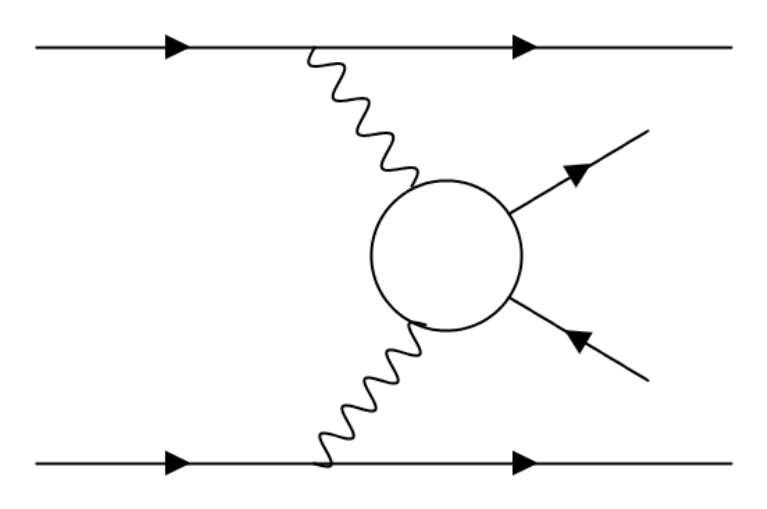}
            \put(20,10){\textcolor{black}{${}^{208}_{82}Pb$}}
            \put(20,54){\textcolor{black}{${}^{208}_{82}Pb$}}
            \put(43,17){\textcolor{black}{$\gamma$}}
            \put(43,47){\textcolor{black}{$\gamma$}}
            \put(51.3,32){\textcolor{black}{SM / BSM}}
            \put(65,10){\textcolor{black}{${}^{208}_{82}Pb^{*}$}}
            \put(65,54){\textcolor{black}{${}^{208}_{82}Pb^{*}$}}
            \put(75,38){\textcolor{black}{$P_{a}$}}
            \put(75,26){\textcolor{black}{$\overline{P_{b}}$}}
            
        \end{overpic}
        \caption{Typical ultra-peripheral collision process of lead ions, where $P_{a}$ and $P_{b}$ represent possible final states.}
        \label{feynmanupc}
    \end{minipage}
    \hfill
\end{figure}

In the particular case where the beam is composed by lead ions, we can illustrate this interaction through Fig. \ref{feynmanupc}  \newline

Here, ${}^{208}_{82}Pb$ (${}^{208}_{82}Pb^{*}$) represents the lead ion before (after) the collision, the SM/BSM box denotes both SM and potential BSM interactions that could occur after the photon-photon induced processes, and $P_{a}$/$P_{b}$ represents the set of particles produced as a result of UPCs that can be detected and analyzed. The coherent emission of photons by a nucleus requires that the wavelength of the photons be larger than the size of the nucleus. This condition ensures that the photons do not probe the internal structure of the hadrons but instead reflect the overall interaction among their components. In this coherent emission process, the photons are constrained to be nearly on the mass shell, enabling them to directly participate in the interaction.  \newline

The physics of heavy ion collisions has been studied since the early days of modern particle physics and has recently gained prominence due to technological advancements in particle colliders, particularly in terms of energy and luminosity \cite{Bruce:2722753, deFavereaudeJeneret:2009db}. Consequently, UPCs offer a cleaner experimental environment, as the processes involved are well-defined by electromagnetic interactions, facilitating the study of specific phenomena related to photon-photon induced processes. In particular, UPCs open a window for exploring BSM theories \cite{Klasen:2008ja, ATLAS:2017fur,Bruce:2018yzs, Klein:2020nvu, dEnterria:2022sut}. For instance, within the context of hadron colliders, studies have investigated a wide range of phenomena, including tau anomalous magnetic moment \cite{delAguila:1991rm,Beresford:2019gww,Dyndal:2020yen}, axion-like particles \cite{Goncalves:2021pdc,Knapen:2016moh,Barbosa:2025zyn}, extra dimensions \cite{Atag:2010bh}, new doubly charged scalars \cite{Babu:2016rcr}, supersymmetric particles \cite{deFavereaudeJeneret:2009db,Godunov:2019jib}, dark matter \cite{Beresford:2018pbt, Harland-Lang:2018hmi}, monopoles \cite{Baines:2018ltl,Dougall:2007tt}, and gravitons \cite{Zhou:2007wfr,İnan_2012}.  \newline

As mentioned above, the primary objective of the first part of this work is to demonstrate that, with an integrated luminosity of $3.48 \text{ nb}^{-1}$ at the ATLAS experiment using PbPb-UPC, it is possible to observe signals of a new charged vector boson in the context of the Vector Scotogenic Model. We specifically focus on the decay of these particles into muons and missing transverse energy (MET), utilizing an efficient selection cut strategy. To this end, the main characteristics of the model are presented in the following section.  \newline

\section{The Vector Scotogenic Model}
\label{vsmmodel}

Among the many proposed SM extensions to simultaneously resolve the origin of neutrino mass and the nature of dark matter, the Vector Scotogenic Model stands out for its theoretical simplicity. Initially, the model was developed for scalar and fermion fields in order to solve the dark matter problem, but it was later demonstrated that this approach works remarkably well for vector fields as well, where the inclusion of massive vector fields that transform as doublets under $SU(2)_L$ group provides a viable DM candidate \cite{Cirelli:2005uq, Belyaev:2018xpf, Saez:2018off}. Motivated by these two open problems of the SM and the simplicity of the model, in this work we study an extended version of this VSM where HNLs are introduced. This specific construction introduces an accidental discrete symmetry that stabilizes the dark matter candidate and generates neutrino masses radiatively at the one-loop level \cite{CarcamoHernandez:2018vdj,C:2022nuo,C:2024kds}. This implies that the parameters that define the relic density and neutrino mass structure can be tested by searching for distinctive signatures at colliders. The dilepton+met channel is one of these signatures predicted by the VSM that can be studied with UPCs collisions for constraining the parameter space related to the production of new charged vector bosons $V^{\pm}$. Before discussing the details of the production of new $V^{\pm}$ via lead-lead and proton-proton UPCs , the most important theoretical details of the model are presented here. \newline

The simplicity of the VSM model arises from the introduction of the following vector boson doublet

\begin{equation}
    {V}_{\mu}= \begin{pmatrix}
{V}^{+}_{\mu} \\ {V}^{0}_{\mu}
\end{pmatrix},
\end{equation}

which transform as $(1,2,1/2)$ under the ${SU(3)}_{C} \times {SU(2)}_{L} \times {U(1)}_{Y}$ symmetry group. The dynamics of these new bosons is described by the following lagrangian  \newline

\begin{align}
\mathcal{L} & \supset -\frac{1}{2} \left( {D}_{\mu}{V}_{\nu} - {D}_{\nu}{V}_{\mu} \right)^{\dagger} \left( {D}^{\mu}{V}^{\nu} - {D}^{\nu}{V}^{\mu} \right) + i\frac{g'}{2}{k}_{1}{V}^{\dagger}_{\mu}{B}^{\mu\nu}{V}_{\nu} + ig{k}_{2}{V}^{\dagger}_{\mu}{W}^{\mu\nu}{V}_{\nu} + {M}_{V}^{2}{V}_{\mu}^{\dagger}{V}^{\mu} \nonumber \\
& - \alpha_{2}({V}_{\mu}^{\dagger}{V}^{\mu})({V}_{\nu}^{\dagger}{V}^{\nu}) - \alpha_{3}({V}_{\mu}^{\dagger}{V}^{\nu})({V}_{\nu}^{\dagger}{V}^{\mu}) - {\lambda}_{2}({\Phi}^{\dagger}\Phi)({V}_{\mu}^{\dagger}{V}^{\mu})-{\lambda}_{3}({\Phi}^{\dagger}{V}_{\mu})({V}^{\mu\dagger}\Phi) \nonumber \\
& -\frac{{\lambda}_{4}}{2}\left [ ({\Phi}^{\dagger}{V}_{\mu})({\Phi}^{\dagger}{V}^{\mu})+({V}^{\mu\dagger}\Phi)({V}_{\mu}^{\dagger}\Phi) \right ], \label{vsmlagrangian}
\end{align}

where ${W}_{\mu\nu}$ and ${B}_{\mu\nu}$ are the usual field strength tensors related to the ${SU(2)}_{L}$ and ${U(1)}_{Y}$ groups, respectively, while $\Phi$ is the SM Higgs field. The coupling constants ${k}_{1,2}$, ${\alpha}_{2,3}$ and ${\lambda}_{2,3,4}$ are free parameters. A ${Z}_{2}$ symmetry was assumed here, so the neutral component of ${V}_{\mu}$ may be a dark matter candidate. The first, second and third terms of the lagrangian given by Eq. \ref{vsmlagrangian} describe the interaction between the new field ${V}_{\mu}$ and SM gauge bosons. The fourth, fifth and sixth terms correspond to a mass term for ${V}_{\mu}$ and self-interaction terms. Finally, the last three terms correspond to the interaction of ${V}_{\mu}$ with the Higgs field $\Phi$.  \newline

In this work we are interested in the production of charged vector bosons ${V}^{\pm}$ and its subsequent decay into muons and HNLs. For this, we introduce into the lagrangian of Eq. \ref{vsmlagrangian}, the interaction of ${V}_{\mu}$ with HNLs given by the following expression  \newline

\begin{equation}
    \mathcal{L} = -\sum_{i=e,\mu,\tau}{\beta}_{i}\overline{{L}_{i}}{\gamma}^{\mu}{\widetilde{V}}_{\mu}{N}_{L} + h.c.,
    \label{hnlvlagrangian}
\end{equation}

with

\begin{equation}
    {\widetilde{V}}_{\mu}=i{\sigma}_{2}{V}_{\mu}^{*}=\frac{1}{\sqrt{2}}\begin{pmatrix}
{V}_{\mu}^{1}-i{V}_{\mu}^{2} \\ -\sqrt{2}{V}_{\mu}^{-}
\end{pmatrix} .
\end{equation}

In Eq. \ref{hnlvlagrangian}, ${\beta}_{i}$ are coupling constants and ${N}_{L}$, which acts within this model as a portal between $V$ and the SM leptons ${L}_{i}$, represents the HNL with mass ${M}_{{N}_{L}}$. In this model ${N}_{L}$ is odd under ${Z}_{2}$ symmetry, such that ${N}_{L}$ may also be a candidate for dark matter. The parameters $k_i$ and $\beta_i$ relevant to our analysis were chosen following the analysis performed in \cite{C:2022nuo}. To avoid interactions between the photons and the neutral component of the vector doublet, the parameters $k_i$ are set to $k_1$=$k_2$=1. Additionally, to ensure the scenario is consistent with perturbativity and dark-matter constraints, we choose $\beta_1 = 0$, $\beta_2 = \beta_3 = 0.5$.  \newline

To investigate the production of a new charged vector boson and its subsequent decay into dimuon+MET final states via ultraperipheral collisions, it is essential to first review the current landscape of charged vector boson searches. The hunt for new charged vector bosons has long been a central focus in collider physics; while various experiments have constrained the properties of these hypothetical particles, the most significant mass limits to date have been established in the high-energy regime by the ATLAS and CMS experiments using conventional proton-proton collisions. In particular, the most stringent constraint has been obtained from the $\ell\nu_{\ell}$ channel, where constraints on the production and branching ratios for a new charged vector boson were established across $\sim$ 0.15-7 TeV mass ranges at a center-of-mass energy of 13 TeV and a luminosity of 140 ${fb}^{-1}$. For this channel, ATLAS and CMS collaborations have established the mass limits of the charged vector boson in ${M}_{{W}^{\prime}}>6$ TeV and $>$ 5.7 TeV, respectively \cite{ATLAS:2019lsy, CMS:2022krd}. In the same search but through the $\mu\nu$ channel, CMS established the limit at ${M}_{{W}^{\prime}}>5.6$ TeV \cite{CMS:2022krd}. Similar searches conducted by the ATLAS and CMS collaborations in the ${W}^{\prime} \rightarrow \tau \nu$ channel at $\sqrt{s} = 13$ TeV, with integrated luminosities of 139 ${fb}^{-1}$ and 138 ${fb}^{-1}$ respectively, have probed mass ranges from 500 GeV to 6 TeV. These two searches establish a lower limit on the ${W}^{\prime}$ mass in ${M}_{{W}^{\prime}} > 5$ TeV \cite{ATLAS:2024tzc, CMS:2022ncp}. Other channel explored by ATLAS and CMS include ${W}^{\prime} \rightarrow 2j$, covering the mass range from 1.5 to 8 TeV with integrated luminosities of 139 ${fb}^{-1}$ and 137 ${fb}^{-1}$, respectively, establishing the lower mass limit in 4 TeV \cite{ATLAS:2019fgd, CMS:2019gwf}. Additional mass limits have been reported for semileptonic decays, placing constraints in 3.9 TeV within the context of the Heavy Vector Triplet model \cite{ATLAS:2020fry}. Through the diboson production, other searches have established limits at 4.8 TeV \cite{CMS:2022pjv}. We highlight that these mass constraints for charged vector bosons are model-dependent, and the mass limits showed here were for a model where this new particle has couplings to fermions identical to those of the W boson of the SM and the Heavy Vector Triplet model. Furthermore, it is important to emphasize that these limits were obtained through conventional proton-proton collisions\footnote{For a long time, constraints on $M_{{V}^{\pm}}$ could be obtained from discrepancies between theory and data from the muon anomalous magnetic moment. In this work we will not take into account these constraints, since the recent Fermilab measurement shows that the experimental value is close to that predicted by the SM \cite{Muong-2:2025xyk}.}. This means that in low-mass regions, the sensitivity of conventional searches can be affected because the background from QCD processes is significantly high. In this mass region, UPC collisions can complement conventional searches since, as mentioned, in UPCs the electromagnetic interaction is dominant and does not produce considerable QCD background. Constraints in the low-mass region are often indirectly inferred from slepton (spin-0) searches conducted at LEP-II \cite{lepsusy2001}. However, it is crucial to clarify that LEP-II did not perform dedicated searches for new charged vector bosons (spin-1) within this mass range. While the large theoretical cross-section for ${e}^{+}{e}^{-}\rightarrow{\gamma}^{*}\rightarrow {W}^{\prime +}{W}^{\prime -}$ suggests that masses $M_{V^\pm} < 105$ GeV might be excluded \cite{ParticleDataGroup:2024cfk,ALEPH:2002nwp, lepsusy2001}, these mass limits are highly model-dependent and assume specific coupling structures and decay kinematics optimized for scalars particles. Consequently, this mass region remains a viable and compelling window for phenomenological exploration via Ultra-Peripheral Collisions, where the production mechanism and backgrounds differ significantly from traditional searches. In summary, in the following sections, we propose a direct search strategy for these spin-1 charged bosons to provide a direct and definitive experimental test, instead of relying on indirect evidence\footnote{To maintain clarity, we refer to this mass range in the results section as the 'Indirect Exclusion by LEP-II', distinguishing between reinterpreted slepton searches and the direct search of our proposed PbPb-UPC study.}.\newline

Therefore, given the lack of direct searches for new vector bosons in the $M_{W^{\prime}} < 105$ GeV range, we will demonstrate that the ATLAS experiment can probe this mass window using ultra-peripheral Pb-Pb collisions at $\sqrt{{s}_{NN}} = 5.02$ TeV. Utilizing an integrated luminosity of $\mathcal{L}_{int} = 3.48$ ${nb}^{-1}$, we show that this channel offers sufficient sensitivity to explore favorable regions of the VSM parameter space\footnote{The analysis presented in the first part of this work focuses on the ATLAS experiment. However, the methodology used to derive these estimates is equally applicable to the CMS experiment. While there are subtle differences between the two experiments, these can be addressed in subsequent studies.}. In the next section, we present the details of the analysis of the production of new charged vector bosons and their subsequent decay into dimuon+MET at the ATLAS experiment.  \newline

\section{Probing New Physics at ATLAS using Ultra-peripheral Collisions of Lead Ions}
\label{upc_atlas}

\subsection{Details of the analysis}
\label{analysis}

As previously mentioned, our goal in this section is to demonstrate that ATLAS Experiment, with PbPb-UPCs at $\sqrt{{s}_{NN}}$ = 5.02 TeV and an integrated luminosity of $\mathcal{L}_{int} = 3.48$ ${nb}^{-1}$, can directly probe signals of new physics arising from the production of a new charged vector boson and its subsequent decay into HNLs predicted by the VSM model. For this purpose, we study the production of these particles and identify the parameter space that favors the signal over the background in UPC events through the production channel ${}^{208}_{82}Pb+{}^{208}_{82}Pb \rightarrow {\mu}^{-} + {\mu}^{+} + MET + {}^{208}_{82}Pb^{*} + {}^{208}_{82}Pb^{*}$, where the MET is attributed to either SM neutrinos or HNLs. For this, we implement the lagrangians of the VSM model given by the Eqs. \ref{vsmlagrangian} and \ref{hnlvlagrangian} and produce the UFO files through the \textit{FeynRules} package. These UFO files are subsequently used by \textit{MadGraph5\_aMC@NLO} package to generate all the signal and background UPC pseudo-events \cite{Degrande:2011ua, Alloul:2013bka, Alwall:2014hca, Shao:2022cly}. For these typical exclusive $\gamma \gamma$ collision processes in UPCs, the probability distribution function \textit{chff} \cite{Vidovic:1992ik} was used.  \newline

For each of the ($M_{{V}^{\pm}}$, $M_{{N}_{L}}$) mass scenarios that we will present throughout this section, 100k UPC events were generated for signal and background at $\sqrt{{s}_{NN}}$ = 5.02 TeV nucleon center-of-mass energy\footnote{The nucleon center-of-mass energy and integrated luminosity assumed here may be exceeded by the LHC by the end of Run 3. Consequently, the results presented in this section likely underestimate the experiment's actual achievable sensitivity.}. These events were initially generated with the following basic selection cuts:  \newline

\begin{align}
& {p}_{T}({\mu}^{\pm}) > 4 \ GeV \nonumber \\
& |{\eta}({\mu}^{\pm})| < 2.4 \nonumber \\
&|{\overrightarrow{ E}}_{T}|_{miss} > 15 \ GeV, \label{selectioncriteria}
\end{align}

where ${p}_{T}({\mu}^{\pm})$ and ${\eta}({\mu}^{\pm})$ represent the transverse momentum and pseudorapidity of muons/antimuons in the final state, while $| {\overrightarrow{E}}_{T}|_{miss}$ is missing transverse energy. Our goal here is to analyze the kinematic and angular observables for muons/anti-muons in order to demonstrate that the ATLAS experiment with $\mathcal{L}_{int} = 3.48$ ${nb}^{-1}$ has the sensitivity enough to probe the existence of a new charged vector boson decaying into HNLs in mass regions not explored by LEP-II.  \newline

The observables selected for this purpose are those shown in Table \ref{table:variables_analysis} where $\Delta R$ is the angular separation between a pair of muons in the $\eta-\phi$ plane, defined as $\Delta R = \sqrt{ {\Delta\eta}^{2} + {\Delta\phi}^{2} }$; $\alpha$ is the acoplanarity between muons, given by $\alpha=1-|\Delta \phi|/\pi$; $M$ is the invariant mass and cos$\theta$ is the cosine of the angle between muon pairs. To estimate the sensitivity of the ATLAS Experiment with PbPb-UPCs, we implemented a random search algorithm in order to identify optimal regions in parameter space where new physics signals can be excluded with a 95\% C.L. or discovered with 5$\sigma$ of statistical significance.   \newline

\begin{table}[]
\begin{tabular}{|c|c|}
\hline
\textbf{Kinematic Observables} & \textbf{Angular Observables} \\ \hline
${p}_{T}$(${\mu}^{-}$)          & $\Delta$R(${\mu}^{-},{\mu}^{+}$)          \\ \hline
${p}_{T}$(${\mu}^{+}$)          & cos$\theta$(${\mu}^{-},{\mu}^{+}$)          \\ \hline
  $|{\overrightarrow{E}}_{T}|_{miss}$         & $\alpha$(${\mu}^{-},{\mu}^{+}$)          \\ \hline
M(${\mu}^{-},{\mu}^{+}$)          & -        \\ \hline
\end{tabular}
\caption{Kinematic and angular observables considered for the UPC event analysis.}
\label{table:variables_analysis}
\end{table}

The detailed results of this analysis are presented in the following section. \newline

\subsection{Results and discussions}
\label{results_upc_atlas_lead}

\begin{figure*}[h!]
    \centering
    \subfloat[]{
        \includegraphics[width=0.48\textwidth]{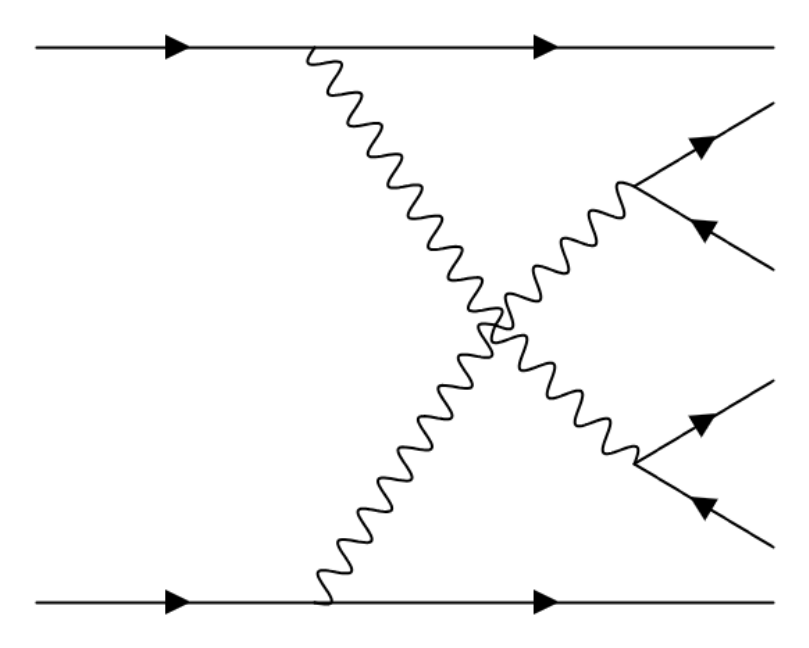}
        \put(-187,150){\textcolor{black}{${}^{208}_{82}Pb$}}
        \put(-85,150){\textcolor{black}{${}^{208}_{82}Pb^{*}$}}
        \put(-187,25){\textcolor{black}{${}^{208}_{82}Pb$}}
        \put(-85,25){\textcolor{black}{${}^{208}_{82}Pb^{*}$}}
        \put(-125,120){\textcolor{black}{$\gamma$}}
        \put(-125,55){\textcolor{black}{$\gamma$}}
        \put(-80,115){\textcolor{black}{${W}^{+}$}}
        \put(-80,55){\textcolor{black}{${W}^{-}$}}
        \put(-40,150){\textcolor{black}{${\nu}_{\mu}$}}
        \put(-40,70){\textcolor{black}{${\mu}^{-}$}}
        \put(-40,105){\textcolor{black}{${\mu}^{+}$}}
        \put(-40,30){\textcolor{black}{$\overline{{{\nu}}}_{\mu}$}}
    }
    \hfill
    \subfloat[]{
        \includegraphics[width=0.48\textwidth]{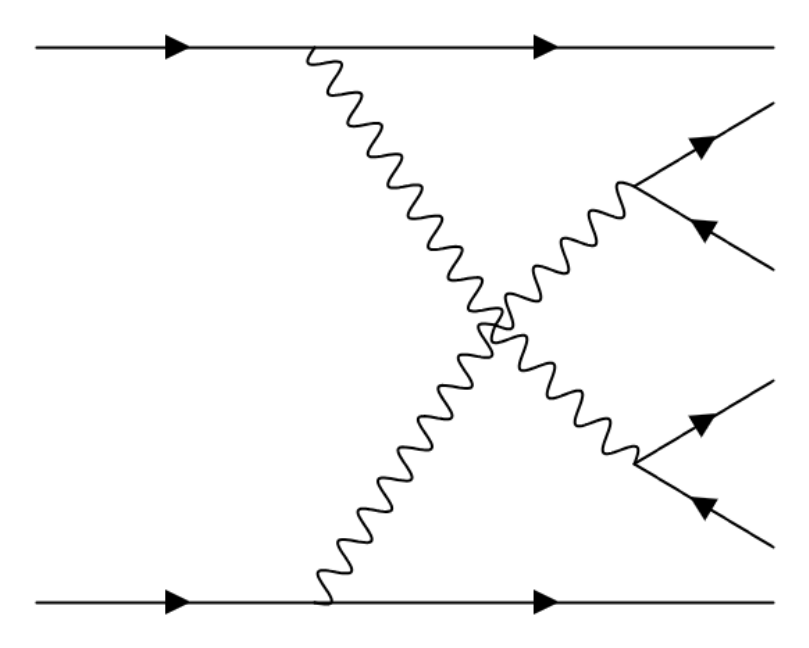}
        \put(-187,150){\textcolor{black}{${}^{208}_{82}Pb$}}
        \put(-85,150){\textcolor{black}{${}^{208}_{82}Pb^{*}$}}
        \put(-187,25){\textcolor{black}{${}^{208}_{82}Pb$}}
        \put(-85,25){\textcolor{black}{${}^{208}_{82}Pb^{*}$}}
        \put(-125,120){\textcolor{black}{$\gamma$}}
        \put(-125,55){\textcolor{black}{$\gamma$}}
        \put(-80,115){\textcolor{black}{${V}^{+}$}}
        \put(-80,55){\textcolor{black}{${V}^{-}$}}
        \put(-40,150){\textcolor{black}{${N}_{L}$}}
        \put(-40,70){\textcolor{black}{${\mu}^{-}$}}
        \put(-40,105){\textcolor{black}{${\mu}^{+}$}}
        \put(-40,30){\textcolor{black}{${N}_{L}$}}
    }
    \caption{Main Feynman diagram contributions for the process ${}^{208}_{82}Pb+{}^{208}_{82}Pb \rightarrow {\mu}^{-} + {\mu}^{+} + MET + {}^{208}_{82}Pb^{*} + {}^{208}_{82}Pb^{*}$ in the (a) background and (b) signal cases.}
    \label{fig:atlas_main_diagrams_lead}
\end{figure*}

The main production mechanism in UPCs contributing to the dimuon+MET final state arises from Vector Boson Fusion (VBF) processes, as represented in Figure \ref{fig:atlas_main_diagrams_lead}.   \newline

\begin{figure*}[h!]
    \centering
    \subfloat[]{
        \includegraphics[width=0.48\textwidth]{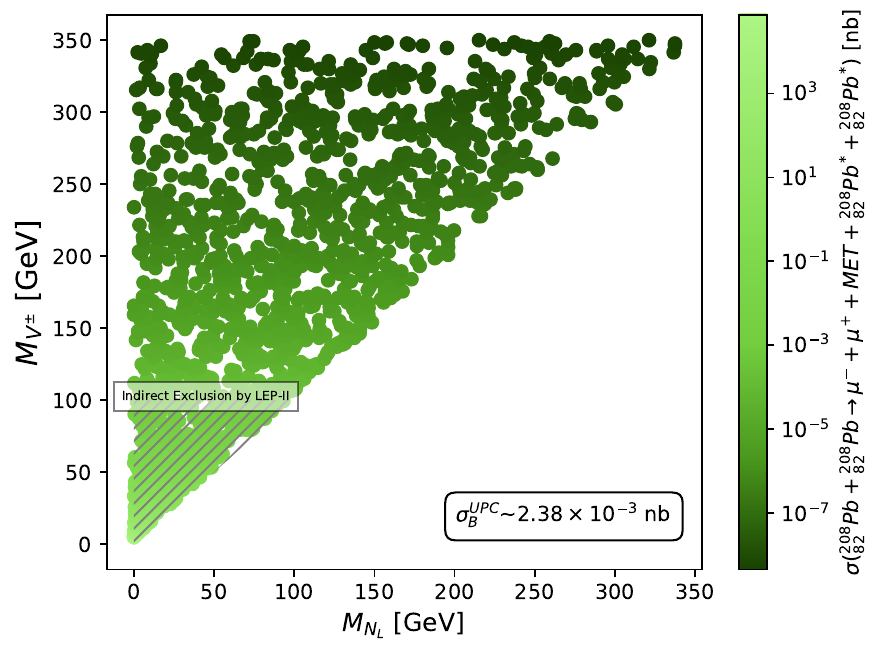}
        \label{fig:lead_cross_section_comparation1}
    }
    \hfill
    \subfloat[]{
        \includegraphics[width=0.48\textwidth]{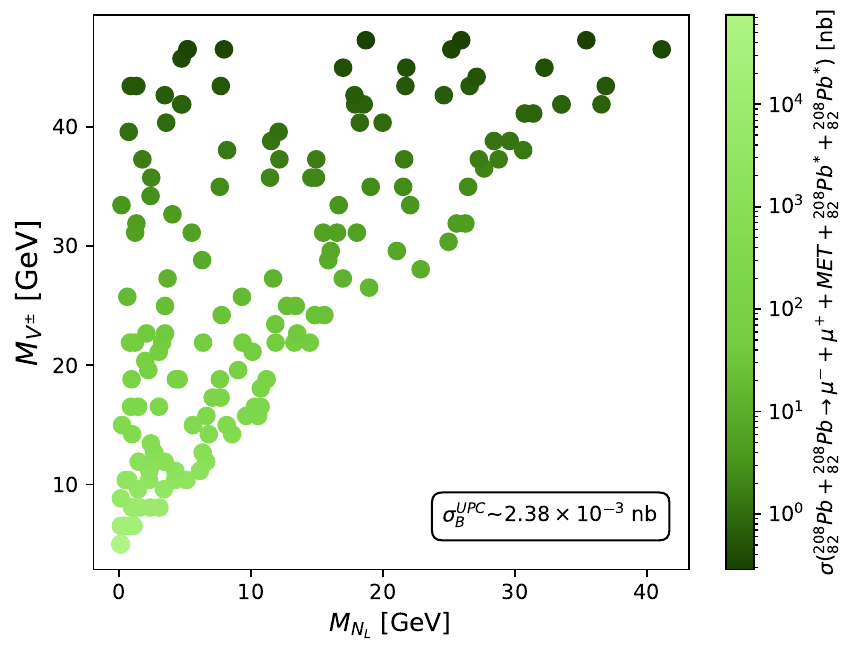}
        \label{fig:lead_cross_section_comparation2}
    }
    \caption{Cross-section as a function of (${M}_{{V}^{\pm}}$, ${M}_{{N}_{L}}$) with ultra-peripheral collisions of lead ions at $\sqrt{{s}_{NN}}$ = 5.02 TeV, where (a) shows the cross-section being compared with the exclusion region suggested by LEP-II, while (b) shows the cross-section points where $ {N}_{S} \geq 1$.}
    \label{fig:crosssection_upc_atlas_lead}
\end{figure*}

The result of the cross-section calculation for dimuon+MET production via V± decays in PbPb-UPCs is shown in Fig. \ref{fig:crosssection_upc_atlas_lead}. For this analysis, we adopt the value ${\beta}_{2} = 0.5$ for the coupling constant between these vector bosons and the SM leptons (see Eq. \ref{hnlvlagrangian}) and initially select UPC events that satisfy the basic cuts listed in Eq. \ref{selectioncriteria}. We assume a muon detection efficiency of 0.8, to be conservative. In Fig. \ref{fig:lead_cross_section_comparation1}, the signal cross-section values for different (${M}_{{V}^{\pm}}$, ${M}_{{N}_{L}}$) mass scenarios in the 5 GeV $< M_{V^{\pm}}, M_{N_L}<$ 350 GeV mass range are compared with the the constraint suggested by LEP-II\footnote{The selection of this mass range is primarily motivated by the effective photon-photon center-of-mass energy accessible in ultra-peripheral lead-ion collisions at the LHC, where the distribution peaks near $\sqrt{{s}^{max}_{\gamma\gamma}} \sim 160$ GeV \cite{Bruce:2018yzs,Klein:2020nvu,dEnterria:2022sut}.}.  \newline

In the other hand, for $\mathcal{L}_{int}$=3.48 ${nb}^{-1}$, only the mass regions depicted in Fig. \ref{fig:lead_cross_section_comparation2} are predicted to yield at least one observable signal event. Consequently and for simplicity, the mass region chosen to estimate the sensitivity at ATLAS experiment is 5 GeV $<{M}_{{V}^{\pm}}$,${M}_{{N}_{L}}<$ 50 GeV, falling in the region where LEP-II did not perform the search.   \newline 

\begin{figure*}[h!]
    \centering
    \subfloat[]{
        \includegraphics[width=0.48\textwidth]{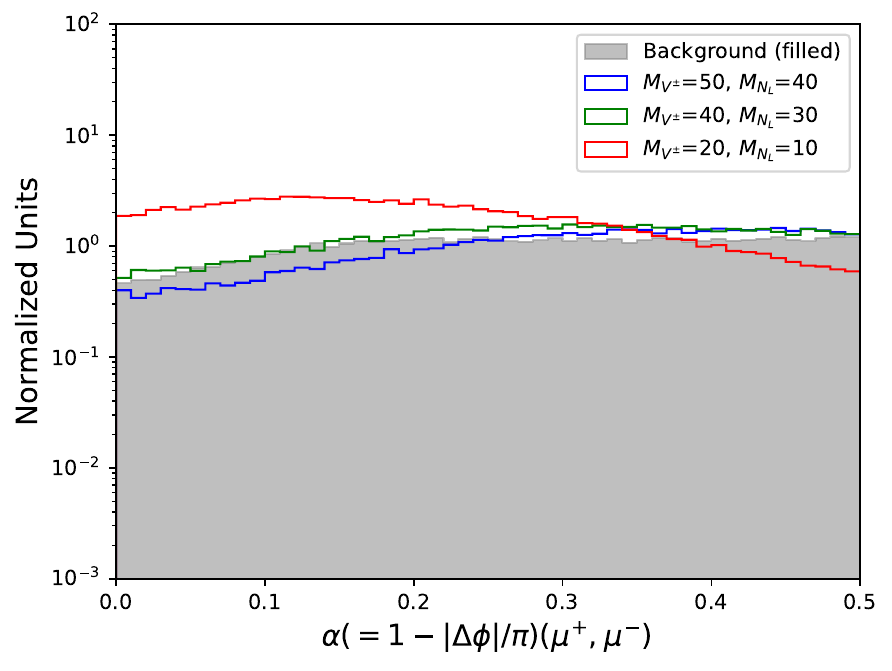}
        \label{fig:acoplanarity_lead}
    }
    \hfill
    \subfloat[]{
        \includegraphics[width=0.48\textwidth]{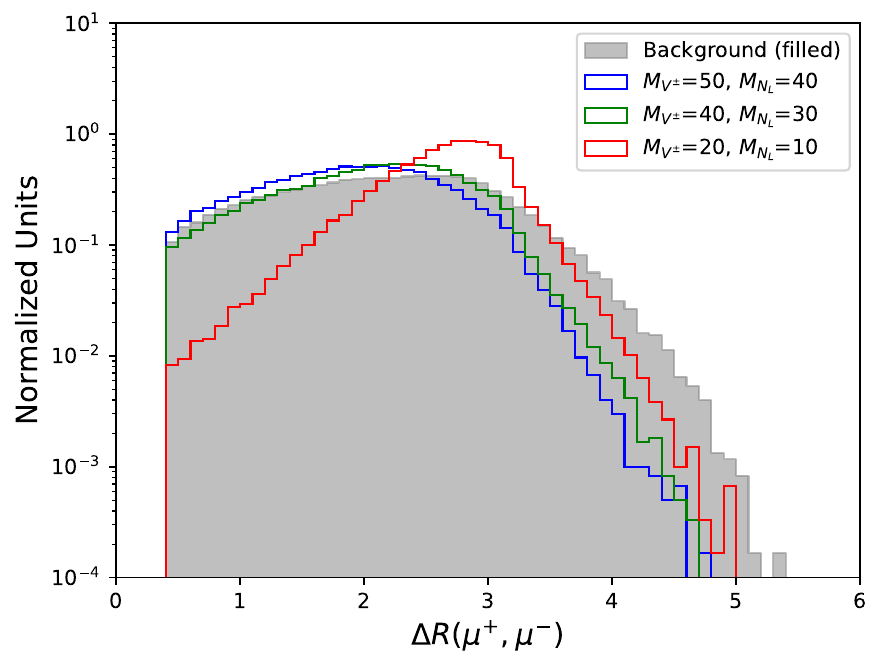}
        \label{fig:delta_r_separation_lhc_05_lead}
    }
    \label{fig:mainfig_lead}
    \vspace{0.5cm} 
    \subfloat[]{
        \includegraphics[width=0.48\textwidth]{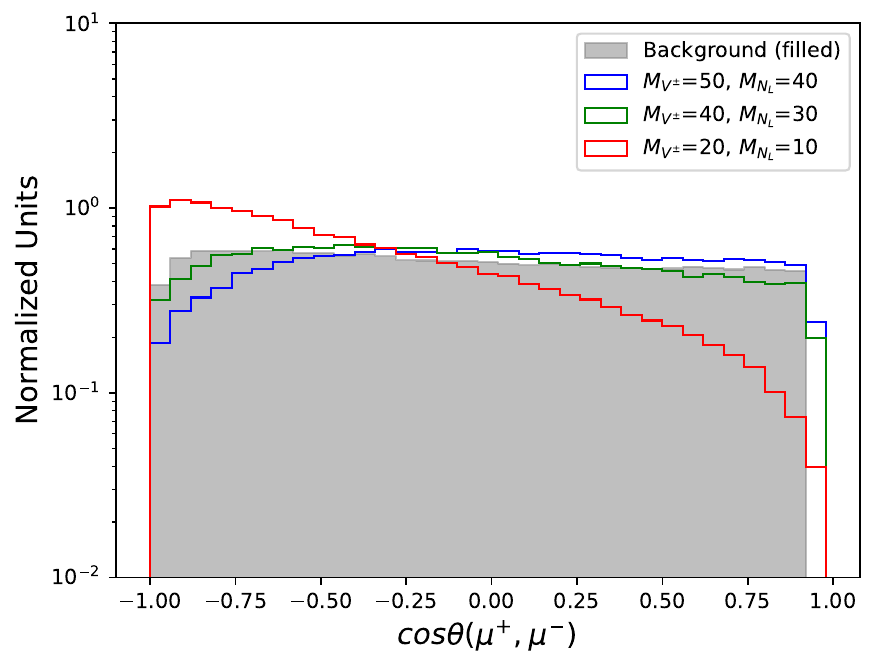}
        \label{fig:costheta_mu1_mu2_lead}
    }
    \caption{Angular distributions for three mass scenarios and the background in PbPb-UPCs at the LHC with $\sqrt{s_{NN}} = 5.02$ TeV, where (a) is the acoplanarity, (b) is the separation in the $\eta \times \phi$ plane, and (c) is the cosine of the separation angle. The mass values are given in GeV.}
    \label{fig:angular_lead}
\end{figure*}

Figure \ref{fig:angular_lead} compares the background distribution with three mass scenarios, showing the acoplanarity $\alpha$, angular separation $\Delta R$, and cos$\theta$ distributions for pairs of hard muons.  \newline 

The shape of the angular distributions shown in Fig. \ref{fig:angular_lead} exhibits a strong dependence on the $(M_{V^{\pm}}, M_{N_{L}})$ masses, which arises from the production and decay kinematics of $V^{\pm}$. Lighter $V^{\pm}$ are typically produced boosted, so most events are concentrated around $\Delta\phi \sim \pi$ (where $\alpha \sim 0$). This also accounts for the $\Delta R$ distributions, since $\Delta\phi \sim \pi$ implies $\Delta R \sim \pi$. In contrast, larger masses tend to be produced at rest, resulting in broader distributions that more closely resemble the background. Similarly, the cos$\theta$ distribution between two pairs of muons show the same analysis. For lighter $V^{\pm}$, an excess of events is observed at $\cos\theta \sim -1$ because the muons from boosted $V^{\pm}$ decays are nearly back-to-back ($\theta_{\mu\mu} \sim \pi$). In the case of larger masses, the lack of boosted $V^{\pm}$ allows the muons to leave in more random directions, resulting in more spreader distributions.   \newline

\begin{figure*}[h!]
    \centering
    \subfloat[]{
        \includegraphics[width=0.48\textwidth]{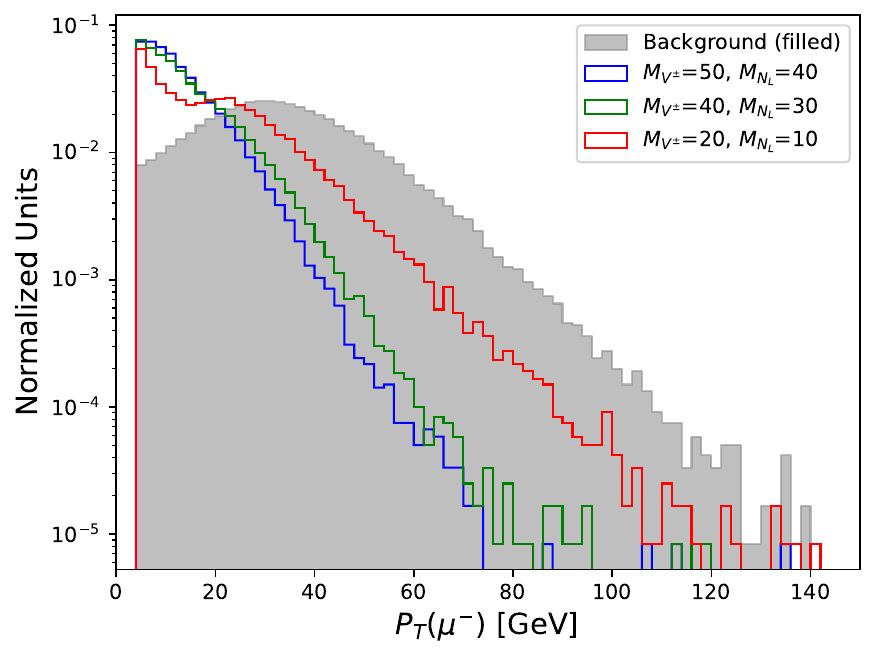}
        \label{fig:muon_pt_05_lead}
    }
    \hfill
    \subfloat[]{
        \includegraphics[width=0.48\textwidth]{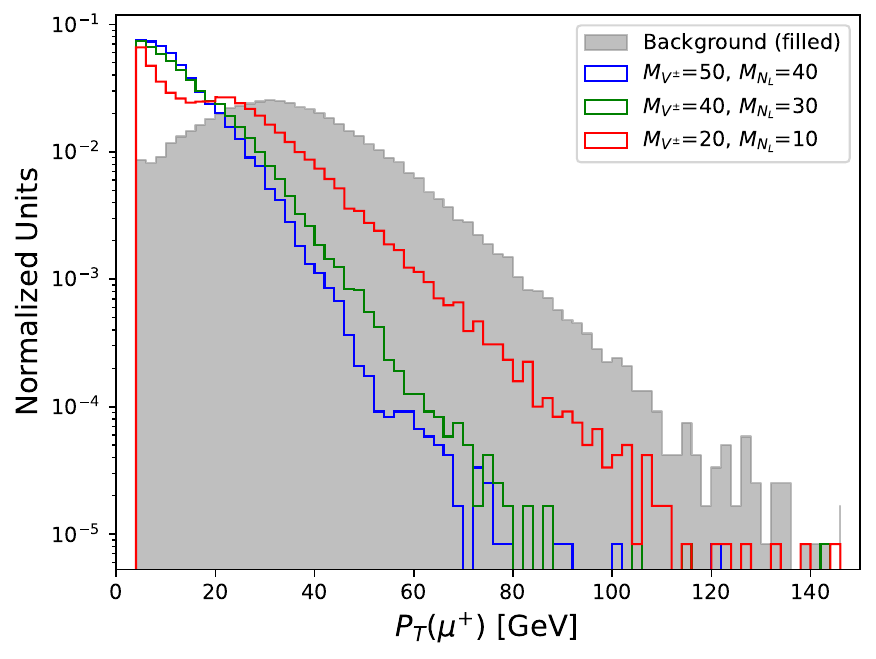}
        \label{fig:antimuon_pt_05_lead}
    }
    \label{fig:mainfig_lead}
    \vspace{0.5cm} 
    \subfloat[]{
        \includegraphics[width=0.48\textwidth]{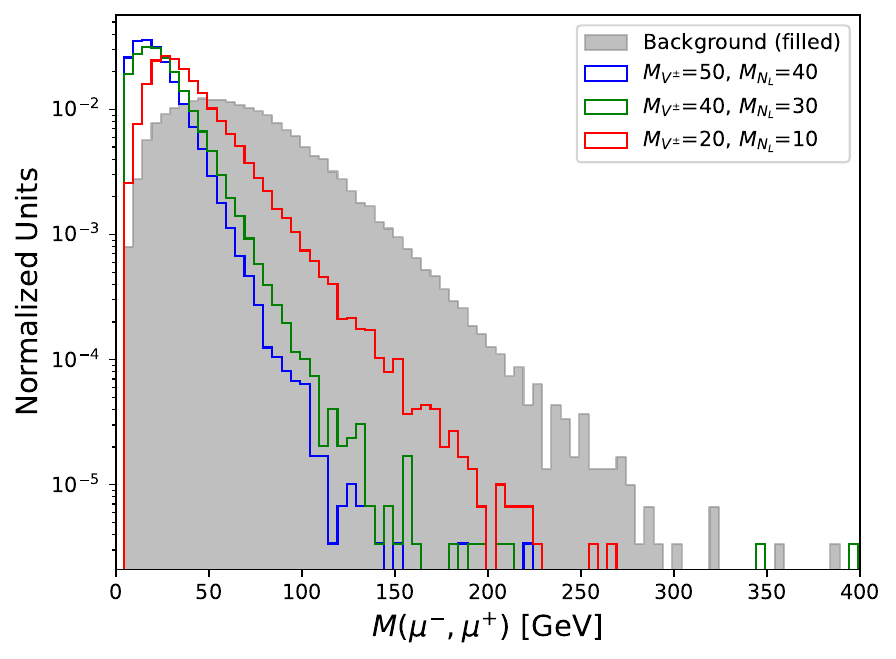}
        \label{fig:invmass_05_lead}
    }
    \hfill
    \subfloat[]{
        \includegraphics[width=0.48\textwidth]{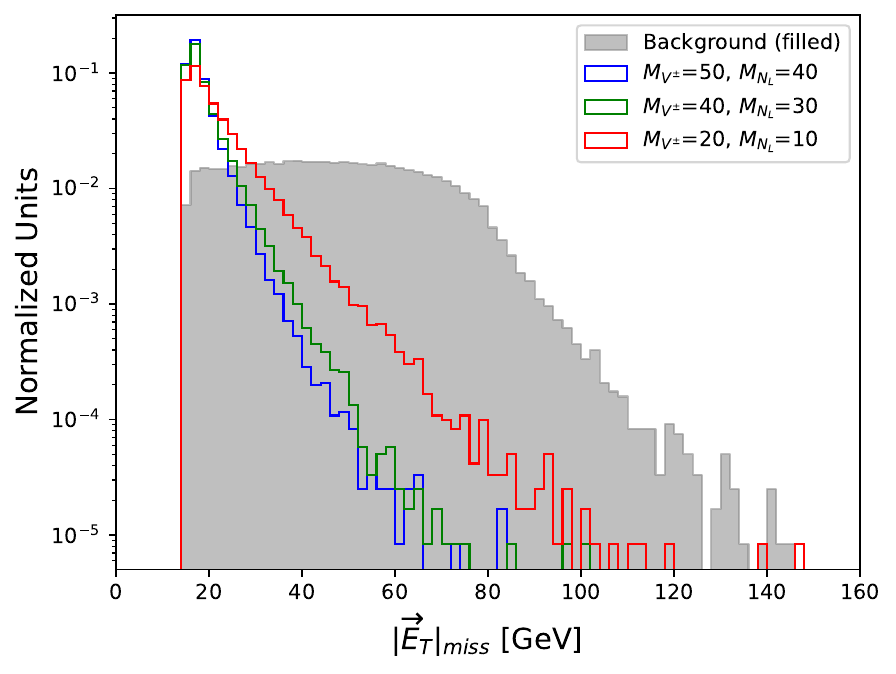}
        \label{fig:met_05_lead}
    }
    \caption{Kinematic distributions for three mass scenarios and the background in PbPb-UPCs at the LHC with $\sqrt{s_{NN}} = 5.02$ TeV, where (a) is the muon transverse momentum, (b) is the anti-muon transverse momentum, (c) is the invariant mass of a pair of muons, and (d) is the missing transverse energy. The mass values are given in GeV.}
    \label{fig:kinematic_lead}
\end{figure*}

On the other hand Fig. \ref{fig:kinematic_lead} shows the kinematic distributions for muon/anti-muon transverse momentum ${p}_{T}$, invariant mass $M_{\mu\mu}$ of muon pairs and missing transverse energy, $|{\overrightarrow{ E}}_{T}|_{miss}$. 

The kinematic distributions presented in Fig. \ref{fig:kinematic_lead} reveal distinctive features that differentiate the signal from the background. The transverse momentum distributions for both muons and antimuons (Figs. \ref{fig:muon_pt_05_lead} and \ref{fig:antimuon_pt_05_lead}) show that muons from the signal processes are predominantly softer, peaking below $p_T \approx 20$ GeV, whereas the background one is much harder, extending significantly beyond 60 GeV due to the large $W$ mass. Similarly, the invariant mass $M_{\mu\mu}$ shown in Fig. \ref{fig:invmass_05_lead} exhibits a sharp peak at lower mass values, contrasting with the broader background distribution. The MET distribution in Fig. \ref{fig:met_05_lead} corroborates this behavior. Here the HNLs carry away less momentum than the SM neutrinos, resulting in a significantly softer $|{\overrightarrow{ E}}_{T}|_{miss}$ for the signal, which falls off rapidly compared to background.  \newline

For the UPC event analysis, a random search algorithm was employed to optimize event selection. A total of $2\times{10}^{5}$ random searches were carried out for each (${M}_{{V}^{\pm}}$, ${M}_{{N}_{L}}$) mass scenario within the parameter space defined by the kinematic and angular observables listed in Table \ref{table:variables_analysis}. Through random exploration, this algorithm finds the optimal angular and kinematic cuts that maximize the statistical significance, thereby enhancing signal efficiency while simultaneously suppressing background events. The significance $S$ chosen for this analysis is shown in Eq. \ref{significance}.  In this expression, $\mathcal{L}_{int}$ represents the integrated luminosity of the ATLAS experiment with Pb-Pb collisions, ${\epsilon}_{S}$ (${\epsilon}_{B}$) denotes the signal (background) selection efficiency for the applied cuts, ${\sigma}^{UPC}_{S}$ (${\sigma}^{UPC}_{B}$) is the signal (background) cross-section, and ${\epsilon}^{sys}_{B}$ is the background systematic error, which was conservatively assumed to be 0.1.  \newline

\begin{equation}
     S = \frac{\mathcal{L}_{int} \times {\epsilon}_{S} \times {\sigma}^{UPC}_{S}}{\sqrt{\mathcal{L}_{int} \times {\epsilon}_{B} \times {\sigma}^{UPC}_{B}+{({\epsilon}_{B}^{sys} \times \mathcal{L}_{int} \times {\epsilon}_{B} \times {\sigma}^{UPC}_{B})}^{2}}}.
    \label{significance}
\end{equation}

The regions in the parameter space for different mass scenarios identified by the random search algorithm, which are favorable for the observation of a new charged vector boson decaying to HNLs in the VSM model, are shown in Table \ref{table:table_atlas_cuts_lead}. This table provides the best cuts for the angular and kinematic observables list in Table \ref{table:variables_analysis}, the best selection cut efficiencies achieved after the sequential application of all cuts, and the expected number of signal and background UPC events observable in the ATLAS experiment through PbPb-UPCs.  \newline

\begin{table}[ht]
\begin{center}
\begin{tabular}{|c|c|c|c|c|c|c|c|c|c|c|c|} 
\hline
${M}_{{V}^{\pm}}$ & ${M}_{{N}_{L}}$ & ${{p}_{T}}_{\mu}>$ & $cos\theta_{\mu\mu}$ & $\Delta R_{\mu\mu}$ & $\alpha_{\mu\mu}<$ & $M_{\mu\mu}$ & $|{\overrightarrow{E}}_{T}|_{miss}$ & $N_{s}$ & $\epsilon^{cut}_{s}$ (\%) & $N_{b}$ & $\epsilon^{cut}_{b}$ (\%) \\
\hline
50 & 40 & 4.12 & 0.034$\pm$0.96 & 1.55$\pm$1.40 & 0.999 & 15.92$\pm$13.88 & 24.86$\pm$9.85 & 1 & 75.22 & 1 & 3.90 \\
\hline  
50 & 30 & 4.07 & 0.007$\pm$0.99 & 1.98$\pm$1.60 & 0.992 & 25.65$\pm$24.27 & 41.01$\pm$26.0 & 1 & 75.60 & 1 & 27.7 \\
\hline
50 & 20 & 4.34 & -0.004$\pm$0.99 & 2.32$\pm$1.96 & 0.991 & 39.10$\pm$35.30 & 30.80$\pm$15.7 & 1 & 76.01 & 1 & 28.7 \\
\hline
50 & 10 & 4.06 & -0.007$\pm$0.98 & 2.56$\pm$2.16 & 0.987 & 151.4$\pm$147.9 & 31.8$\pm$16.78 & 1 & 87.40 & 1 & 53.4 \\
\hline
40 & 30 & 4.14 & -0.005$\pm$0.99 & 1.66$\pm$1.61 & 0.981 & 13.25$\pm$11.41 & 16.77$\pm$1.77 & 1 & 40.70 & 1 & 0.70 \\
\hline
40 & 20 & 4.02 & -0.008$\pm$0.99 & 1.88$\pm$1.83 & 0.976 & 23.58$\pm$21.60 & 23.01$\pm$8.01 & 2 & 57.70 & 1 & 6.50 \\
\hline
40 & 10 & 4.04 & -0.002$\pm$0.99 & 2.38$\pm$2.10 & 0.972 & 37.60$\pm$35.40 & 26.62$\pm$11.6 & 2 & 69.53 & 1 & 20.2 \\
\hline
30 & 20 & 4.06 & -0.004$\pm$0.99 & 2.26$\pm$1.20 & 0.560 & 20.69$\pm$12.37 & 17.97$\pm$2.96 & 8 & 33.27 & 1 & 0.98 \\
\hline
30 & 10 & 4.02 & -0.004$\pm$0.99 & 2.12$\pm$1.72 & 0.477 & 23.85$\pm$14.98 & 22.01$\pm$7.01 & 10 & 43.64 & 1 & 4.39 \\
\hline
20 & 10 & 4.04 & -0.340$\pm$0.66 & 2.83$\pm$0.72 & 0.346 & 17.70$\pm$15.60 & 22.80$\pm$7.77 & 57 & 21.90 & 1 & 0.60 \\
\hline
\end{tabular}
\caption{Best angular and kinematic cuts for different (${M}_{{V}^{\pm}}$,${M}_{{N}_{L}}$) scenarios in the Vector Scotogenic Model. The transverse momentum, invariant mass, and missing transverse energy are given in GeV.}
\label{table:table_atlas_cuts_lead}
\end{center}
\end{table}

\begin{figure*}[h!]
    \centering
    \subfloat[]{
        \includegraphics[width=0.48\textwidth]{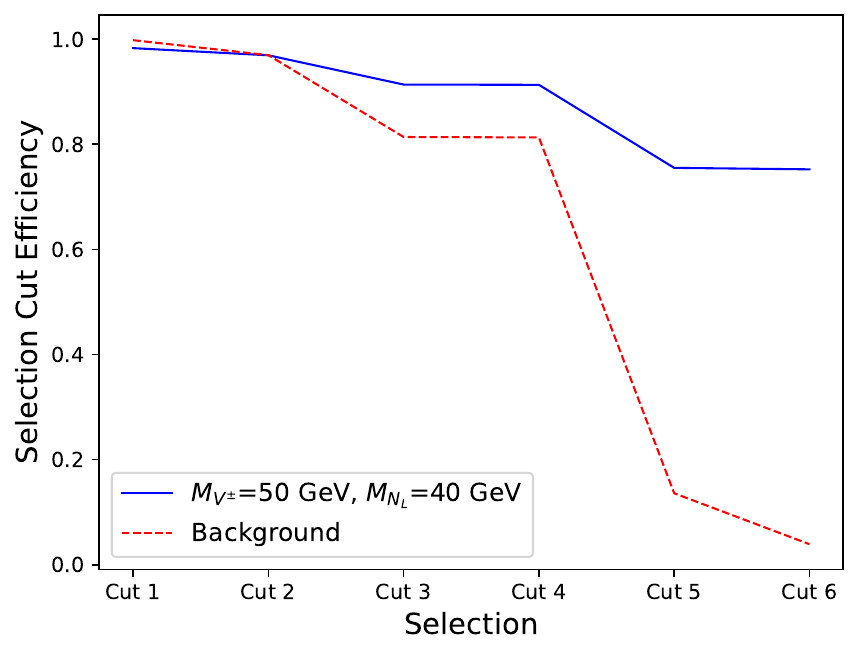}
        \label{fig:cutflow_signal1_lead}
    }
    \hfill
    \subfloat[]{
        \includegraphics[width=0.48\textwidth]{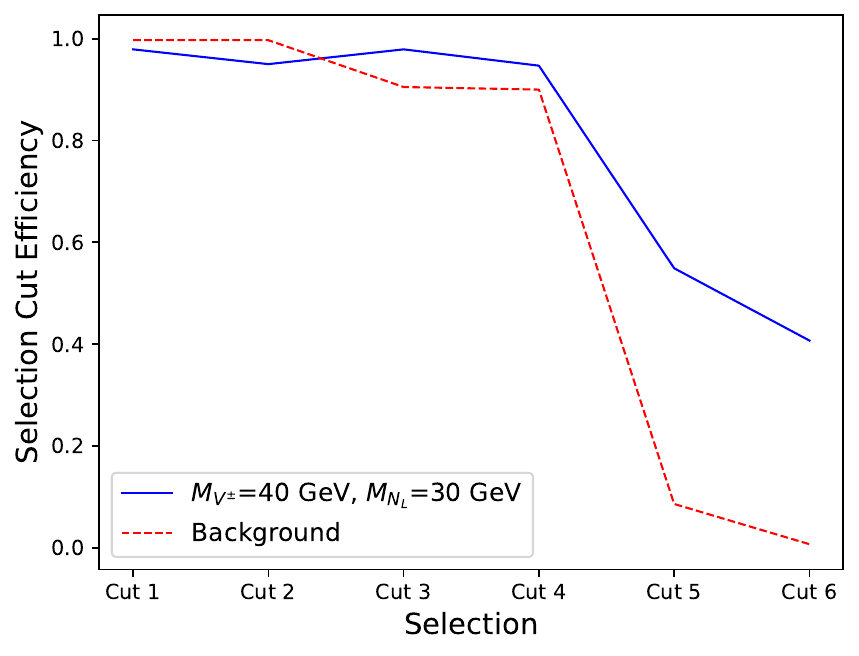}
        \label{fig:cutflow_signal5_lead}
    }
    
    \vspace{0.5cm} 
    \subfloat[]{
        \includegraphics[width=0.48\textwidth]{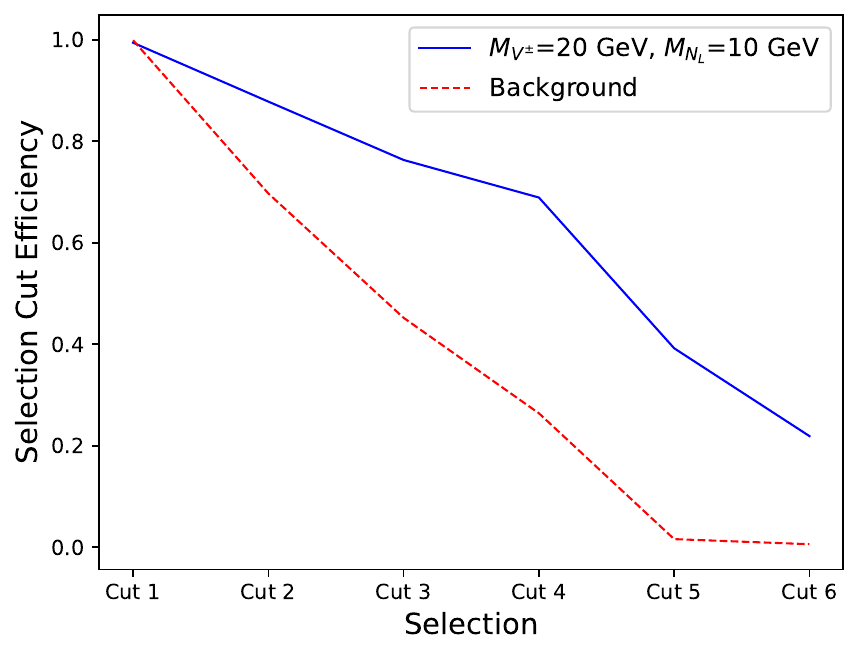}
        \label{fig:cutflow_signal10_lead}
    }
    \caption{Cut Efficiency across each selection cut between the background and three different scenarios: (a) case when ${M}_{{V}^{\pm}} = 50$ GeV, ${M}_{{N}_{L}} = 40$ GeV; (b) case when ${M}_{{V}^{\pm}} = 40$ GeV, ${M}_{{N}_{L}} = 30$ GeV; (c) case when ${M}_{{V}^{\pm}} = 20$ GeV, ${M}_{{N}_{L}} = 10$ GeV. The selection cuts are show in Table \ref{table:table_atlas_cuts_lead}.}
    \label{fig:cutflow_lead}
\end{figure*}

Through this table we can observe that the sequential application of these cuts significantly enhanced the detection sensitivity for new physics, with the lowest mass scenarios emerging as the most promising scenarios. The cutflow efficiency for the background and three different mass scenarios is presented in Figs. \ref{fig:cutflow_signal1_lead}-\ref{fig:cutflow_signal10_lead}, where the y-axis represents the selection cut efficiency value and the x-axis represents the sequential application of the optimized cuts shown in Table \ref{table:table_atlas_cuts_lead}. The substantial reduction in the selection cut efficiency across the cut-flow relative to the high signal event retention is a result of the optimization effectiveness. \newline

\begin{figure*}[h!]
    \centering
    \subfloat[]{
        \includegraphics[width=0.48\textwidth]{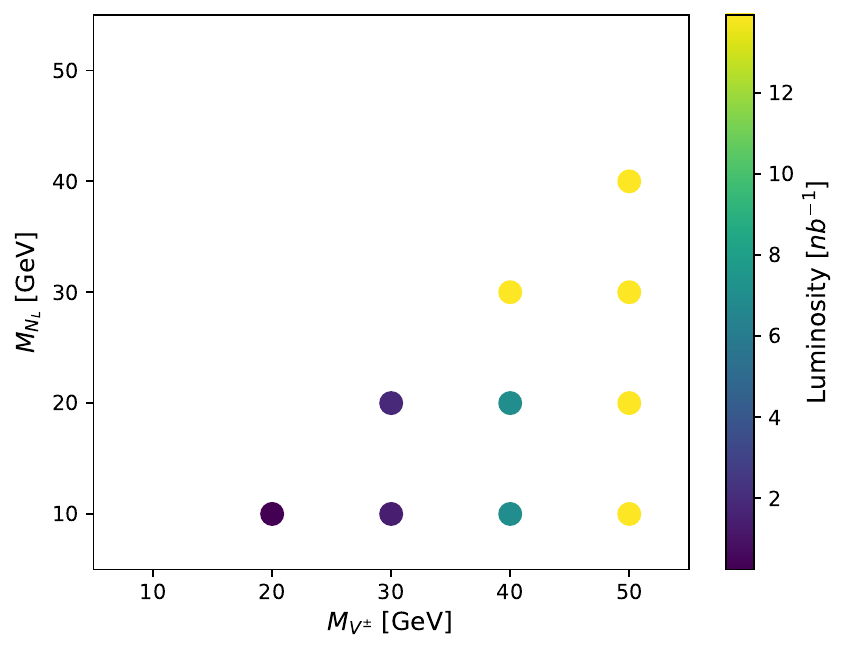}
        \label{fig:atlas_upc_lumi_2sigma_lead}
    }
    \hfill
    \subfloat[]{
        \includegraphics[width=0.48\textwidth]{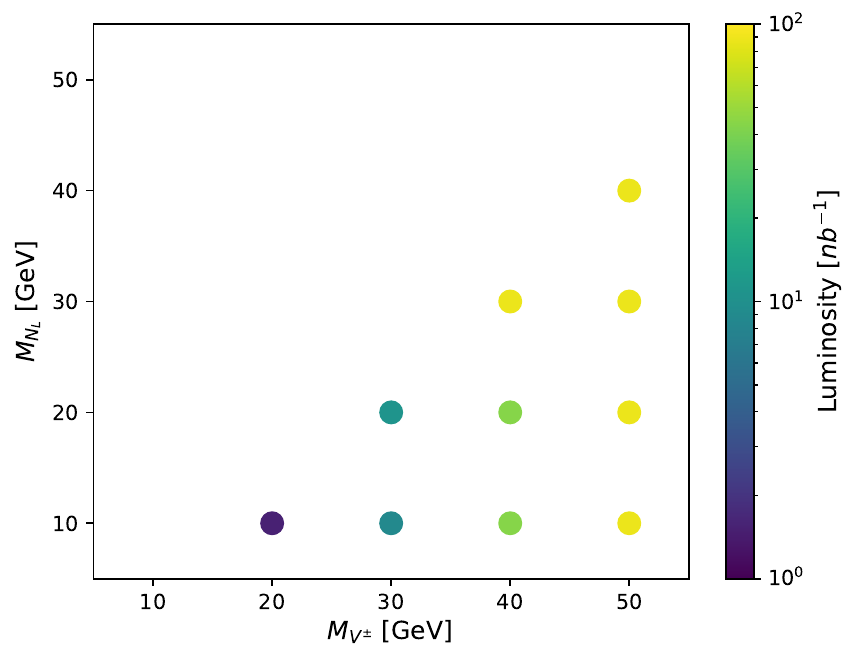}
        \label{fig:atlas_upc_lumi_5sigma_lead}
    }
    
    \caption{Luminosity, in ${nb}^{-1}$, that the ATLAS experiment must reach with UPCs of lead ions (a) to exclude a new charged vector boson at 95\% C.L. and (b) to discover with 5$\sigma$.}
    \label{fig:lumi_atlas_upc_lead}
\end{figure*}

Finally, Fig. \ref{fig:lumi_atlas_upc_lead} show the required luminosity to either exclude a new charged vector boson at 95\% of C.L. (Fig. \ref{fig:atlas_upc_lumi_2sigma_lead}) or to achieve its discovery with 5$\sigma$ (Fig. \ref{fig:atlas_upc_lumi_5sigma_lead}) at ATLAS experiment using ultra-peripheral lead ion collisions. From these figures we can see that for an integrated luminosity $\mathcal{L}_{int}$ = 3.48 ${nb}^{-1}$ currently achieved with ultra-peripheral collisions of lead ions, ATLAS experiment could perform the direct searches for new physics scenarios arising from the decay of a charged vector boson into HNLs at energy scales not covered by previous experiments, such as LEP-II. These figures demonstrate that the ATLAS experiment searching for new charged vector bosons has sufficient sensitivity to exclude or probe different (${M}_{{V}^{\pm}}$, ${M}_{{N}_{L}}$) mass scenarios not explored by previous LEP searches. Specifically, ATLAS with 3.48 ${nb}^{-1}$ can exclude the (30 GeV, 20 GeV), (30 GeV, 10 GeV), and (20 GeV, 10 GeV) scenarios with 95\% CL, and in a discovery scenario, achieve 5$\sigma$ with (20 GeV, 10 GeV).  \newline

This section demonstrates that ultra-peripheral Pb-Pb collisions at the current LHC offer promising sensitivity for direct searches for new charged vector bosons in the $M_{V^{\pm}} < 100$ GeV mass range. While LEP-II did not conduct direct searches for these vector bosons (spin-1), primarily establishing indirect constraints through slepton (spin-0) searches, the current ATLAS experiment, utilizing UPCs of lead ions, does have the potential for discovering these particles at energy scales where conventional proton-proton collisions would produce a significantly higher QCD background. However, despite the signal is cleaner using lead ions, they also introduce inherent challenges, such as background arising from the intricate structure of lead nuclei and the reduced energy per particle compared to lighter beams. Motivated by the advantages offered by ultraperipheral collisions in complementing conventional searches of new particles and considering the higher energy per particle achievable in proton beams compared to lead beams \cite{Bruce:2018yzs,dEnterria:2022sut,Klein:2020nvu}, the following section extends the analysis to higher mass scales at HL-LHC ($\sqrt{s}$ = 14 TeV and $\mathcal{L}_{int}$=150 ${fb}^{-1}$), using ultra-peripheral proton-proton collisions, where we will investigate the sensitivity of the HL-LHC for the $100 < {M}_{{V}^{\pm}} < 600$ GeV mass range and compare our projected results with current existing experimental constrains suggested from SUSY searches.

\section{Probing New Physics at HL-LHC using Ultra-peripheral Proton-Proton Collisions}
\label{upc_14LHC_pp}

Motivated by the high energy and luminosity achievable in proton collisions, this section extends the previous study to the HL-LHC, focusing specifically on pp-UPCs. Before detailing the analysis, a brief comment regarding the background from photon–photon interactions in proton collisions should be made. \newline

We know that a significant fraction of proton collisions at the LHC and HL-LHC can involve photon-photon interactions at different electroweak energies. This is possible thanks to the exchange of photons between very forward-scattered protons \cite{Piotrzkowski:2000rx}, offering another way to probe new charged vector bosons at a larger mass scale. Although it is possible to achieve higher luminosity and energy in these collisions compared to heavy-ion collisions, the different $p^{*}+p^{*}+X$ final states are not free from contamination, particularly from pileup collisions. To mitigate this, two tagging techniques developed for studies of high-energy photon-induced processes in proton collisions are highlighted \cite{deFavereaudeJeneret:2009db}. The first technique requires measuring the very forward-scattered protons in dedicated detectors (VFDs), such as the ATLAS Forward Proton (AFP) detector type. In high-luminosity stages, this approach is mandatory to suppress backgrounds from accidental coincidences, where a proton from a single pileup event is detected in coincidence with a generic proton-proton interaction in the central detector. The second technique, applicable primarily during low-luminosity stages, relies on the presence of Large Rapidity Gaps (LRGs) in the forward directions. This method identifies photon-induced processes by requiring regions of the detector to be devoid of particle activity, effectively suppressing inelastic proton backgrounds. \newline

The availability of these robust tagging methods makes pp-UPCs a particularly compelling environment for the search for new physics. Although the use of tagging techniques and the requirement of intact protons may imply certain limitations on efficiency (such as survival factors and detector acceptance), the massive increase in energy and integrated luminosity compared to heavy ion collisions more than compensates for these effects. Therefore, by combining high energy reach, enhanced luminosity, and robust background control across both low- and high-luminosity operating regimes, pp-UPCs provide a highly promising avenue for discovering new charged vector bosons decaying into HNLs, as will be demonstrated in the following analysis.

\subsection{Details of the analysis}
\label{analysis_upc_14LHC_pp}

\begin{figure*}[h!]
    \centering
    \subfloat[]{
        \includegraphics[width=0.48\textwidth]{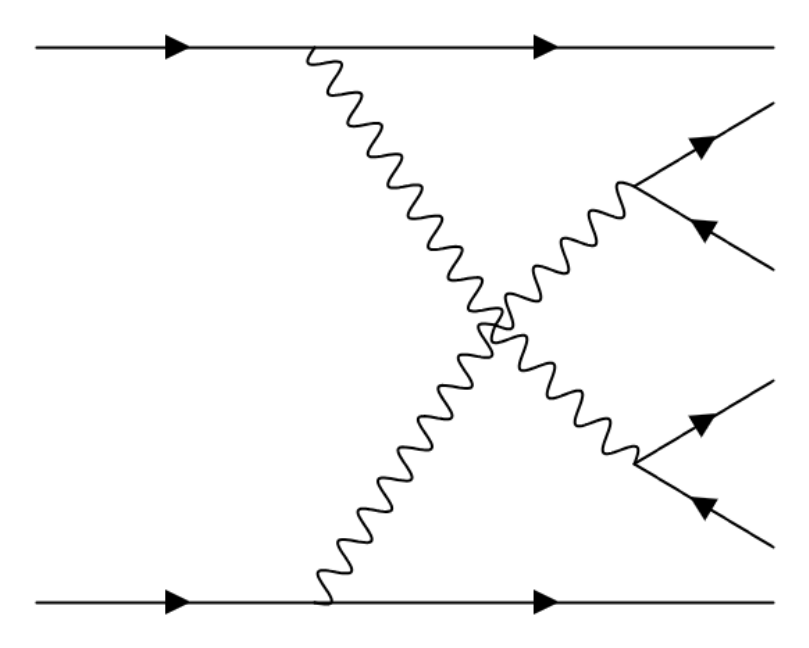}
        \put(-181,152){\textcolor{black}{$p$}}
        \put(-79,152){\textcolor{black}{$p^{*}$}}
        \put(-181,23){\textcolor{black}{$p$}}
        \put(-79,23){\textcolor{black}{$p^{*}$}}
        \put(-125,120){\textcolor{black}{$\gamma$}}
        \put(-125,55){\textcolor{black}{$\gamma$}}
        \put(-80,115){\textcolor{black}{${W}^{+}$}}
        \put(-80,55){\textcolor{black}{${W}^{-}$}}
        \put(-40,150){\textcolor{black}{${\nu}_{\mu}$}}
        \put(-40,70){\textcolor{black}{${\mu}^{-}$}}
        \put(-40,105){\textcolor{black}{${\mu}^{+}$}}
        \put(-40,30){\textcolor{black}{$\overline{{{\nu}}}_{\mu}$}}
    }
    \hfill
    \subfloat[]{
        \includegraphics[width=0.48\textwidth]{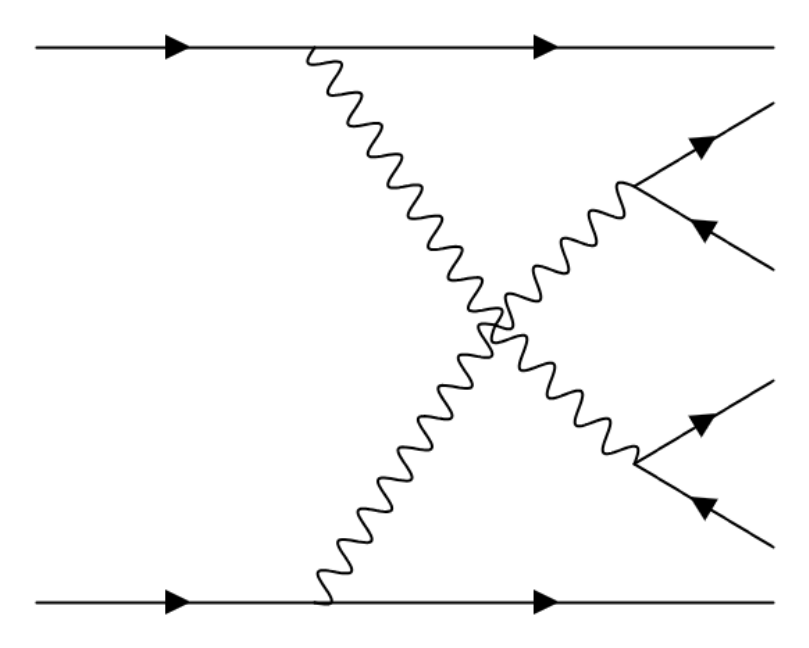}
        \put(-181,152){\textcolor{black}{$p$}}
        \put(-79,152){\textcolor{black}{$p^{*}$}}
        \put(-181,23){\textcolor{black}{$p$}}
        \put(-79,23){\textcolor{black}{$p^{*}$}}
        \put(-125,120){\textcolor{black}{$\gamma$}}
        \put(-125,55){\textcolor{black}{$\gamma$}}
        \put(-80,115){\textcolor{black}{${V}^{+}$}}
        \put(-80,55){\textcolor{black}{${V}^{-}$}}
        \put(-40,150){\textcolor{black}{${N}_{L}$}}
        \put(-40,70){\textcolor{black}{${\mu}^{-}$}}
        \put(-40,105){\textcolor{black}{${\mu}^{+}$}}
        \put(-40,30){\textcolor{black}{${N}_{L}$}}
    }
    \caption{Main Feynman diagram contributions for the process $p+p \rightarrow {\mu}^{-} + {\mu}^{+} + MET + p^{*} + p^{*}$ in the (a) background and (b) signal cases.}
    \label{fig:atlas_main_diagrams_proton}
\end{figure*}

The analysis follows a methodology similar to that used for lead ion collisions. Now we focus on the process $p + p \rightarrow \mu^{-} + \mu^{+} + \text{MET} + p^{*} + p^{*}$, as showed in Fig. \ref{fig:atlas_main_diagrams_proton}.  \newline

As we can see, given the nature of the exclusive $\gamma\gamma$ collision in pp-UPC, the main signal and background contributions are due again to VBF processes. We are interested in investigating the different ($M_{V^{\pm}},M_{N_L}$) mass scenarios that the HL-LHC could exclude at 95\% C.L. or discover with 5$\sigma$ using an integrated luminosity of $ \mathcal{L}_{\text{int}}$ = 150 $ \text{fb}^{-1}$ \cite{Shao:2022cly, Bruce:2018yzs,Klein:2020nvu,dEnterria:2022sut}, excluding or proving the existence of these new particles predicted by the VSM. For this analysis, we have selected $\beta_2=0.5$ and the variables listed in Table \ref{table:variables_analysis} to explore the $5 < {M}_{{V}^{\pm}} < 600$ GeV mass range, now accessible via pp-UPC collisions. For each ($M_{V^{\pm}},M_{N_L}$) mass scenario, we generate 100k signal and background pp-UPC events at a center-of-mass energy of $\sqrt{s} = 14$ TeV with the initial cuts shown in Eq. \ref{selectioncriteria}. Results and discussions are presented in the next section.  \newline

\subsection{Results and discussions}
\label{results_upc_14LHC_pp}

\begin{figure}[h!]
    \centering
    \includegraphics[width=0.6\textwidth]{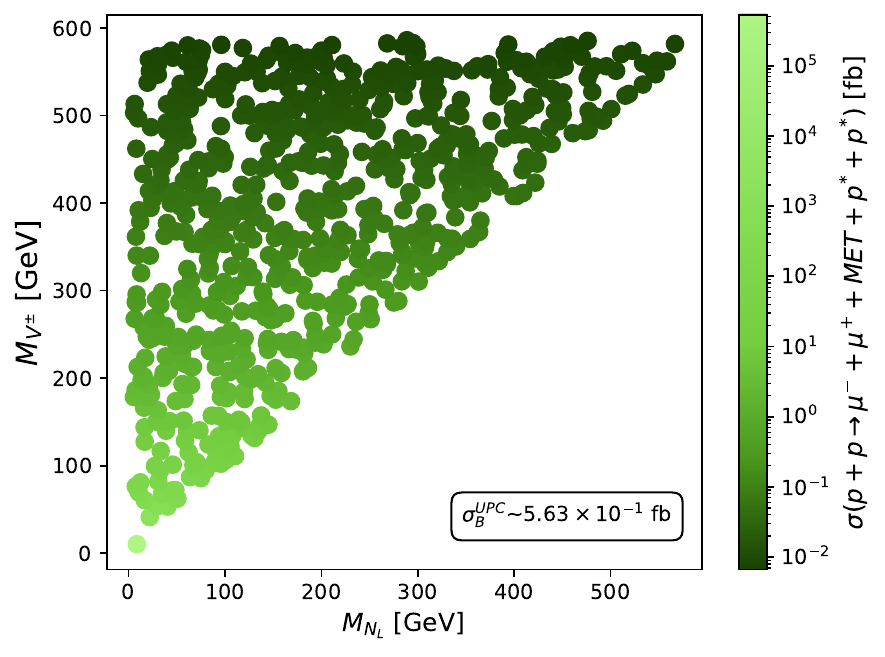}
    \caption{Cross-section as a function of (${M}_{{V}^{\pm}}$, ${M}_{{N}_{L}}$) with ultra-peripheral collisions of protons at $\sqrt{s}$ = 14 TeV.}
    \label{fig:crosssection_upc_atlas}
\end{figure}

Figure \ref{fig:crosssection_upc_atlas} shows the result of the cross-section calculation for different (${M}_{{V}^{\pm}}$,${M}_{{N}_{L}}$) mass scenarios with pp-UPCs at $\sqrt{s}$ = 14 TeV. All cross-section points represent at least one signal event at ATLAS, assuming a muon detector efficiency ${\epsilon}_{eff}$=0.8, to be conservative.  \newline

As illustrated in Figs. \ref{fig:crosssection_upc_atlas_lead} and \ref{fig:crosssection_upc_atlas}, the cross-section values for signal and background UPC processes are higher with Pb-Pb collisions relative to the proton-proton case. This difference is primarily due to the nature of the ultra-peripheral interactions. As previously mentioned, the main characteristic of UPCs is that the nucleus of heavy ions does not fragment during collisions, so the interactions between atomic nuclei are dominated by electromagnetic interactions. In the case of lead ions, which are very heavy, the intense electromagnetic fields generated by the large number of protons per nucleus lead to a wide range of interactions, increasing the cross-section. In contrast, in proton-proton collisions, the absence of these intense electromagnetic fields further suppresses UPC cross-sections. However, the expected luminosity for the HL-LHC with pp-UPCs is significantly higher than the current luminosity achieved by ATLAS with lead ion UPCs \cite{Bruce:2018yzs,Klein:2020nvu,dEnterria:2022sut}. \newline

\begin{figure*}[h!]
    \centering
    \subfloat[]{
        \includegraphics[width=0.48\textwidth]{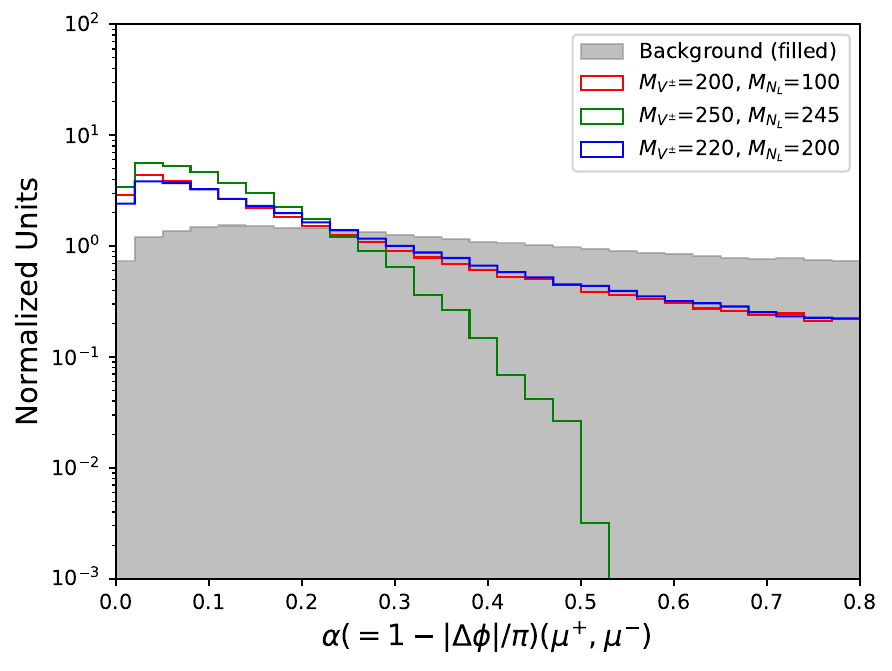}
        \label{fig:acoplanarity}
    }
    \hfill
    \subfloat[]{
        \includegraphics[width=0.48\textwidth]{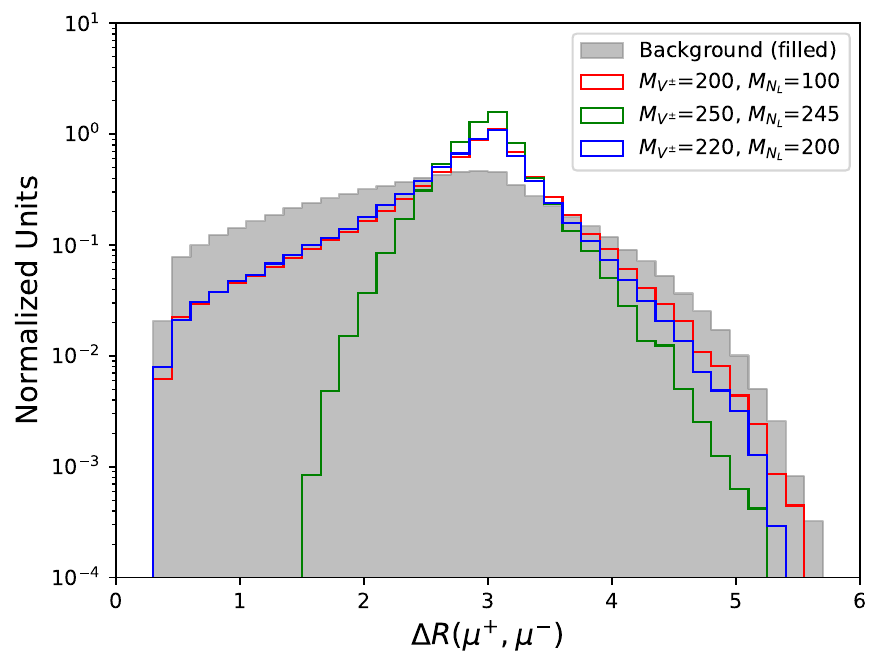}
        \label{fig:delta_r_separation_lhc_05}
    }
    
    \vspace{0.5cm} 
    \subfloat[]{
        \includegraphics[width=0.48\textwidth]{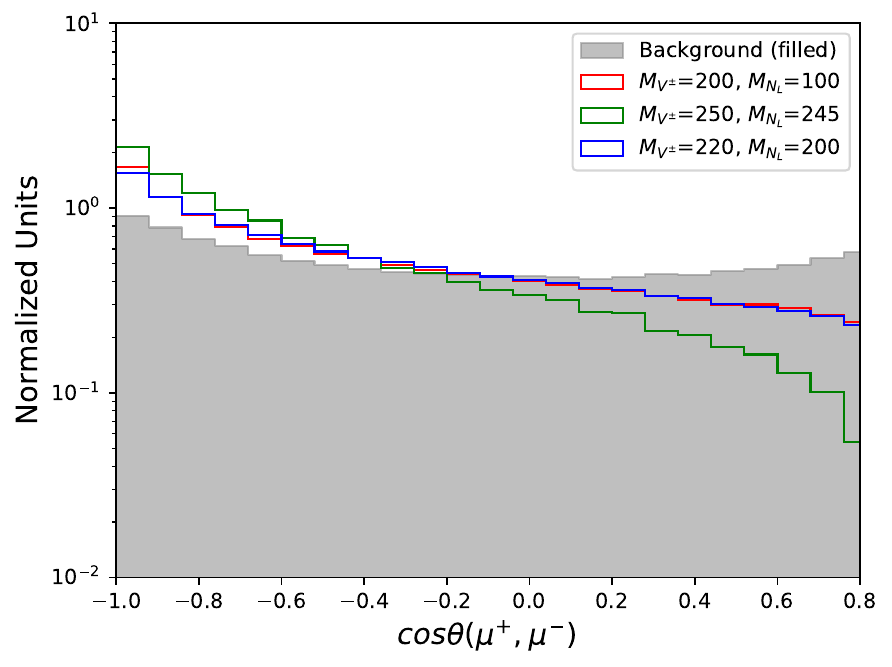}
        \label{fig:costheta_mu1_mu2}
    }
    \caption{Angular distributions for three mass scenarios and the background in pp-UPCs at the HL-LHC with $\sqrt{s}=14$ TeV, where (a) is the acoplanarity, (b) is the separation in the $\eta \times \phi$ plane, and (c) is the cosine of the separation angle. The mass values are given in GeV.}
    \label{fig:angular}
\end{figure*}

\begin{figure*}[h!]
    \centering
    \subfloat[]{
        \includegraphics[width=0.48\textwidth]{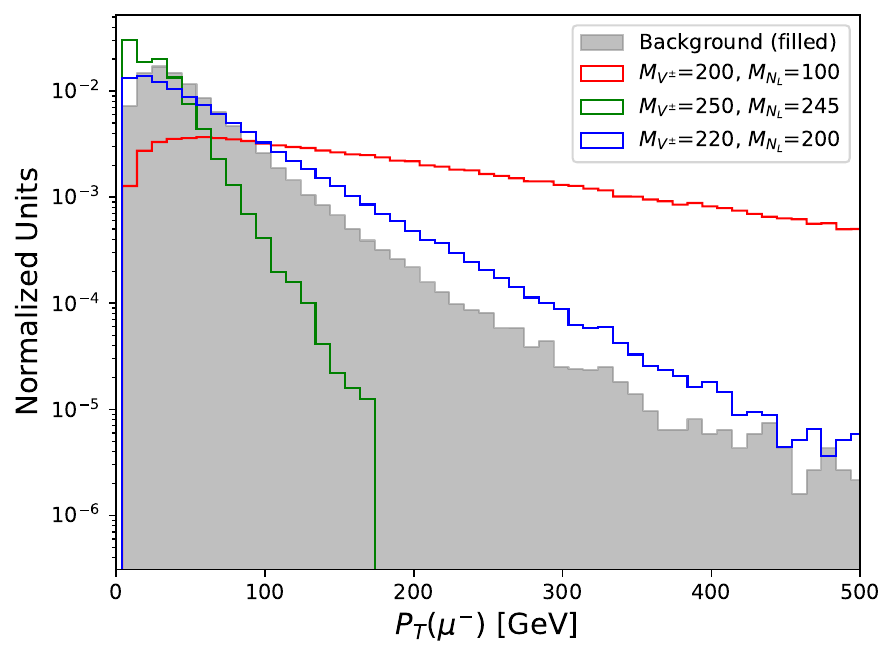}
        \label{fig:muon_pt_05}
    }
    \hfill
    \subfloat[]{
        \includegraphics[width=0.48\textwidth]{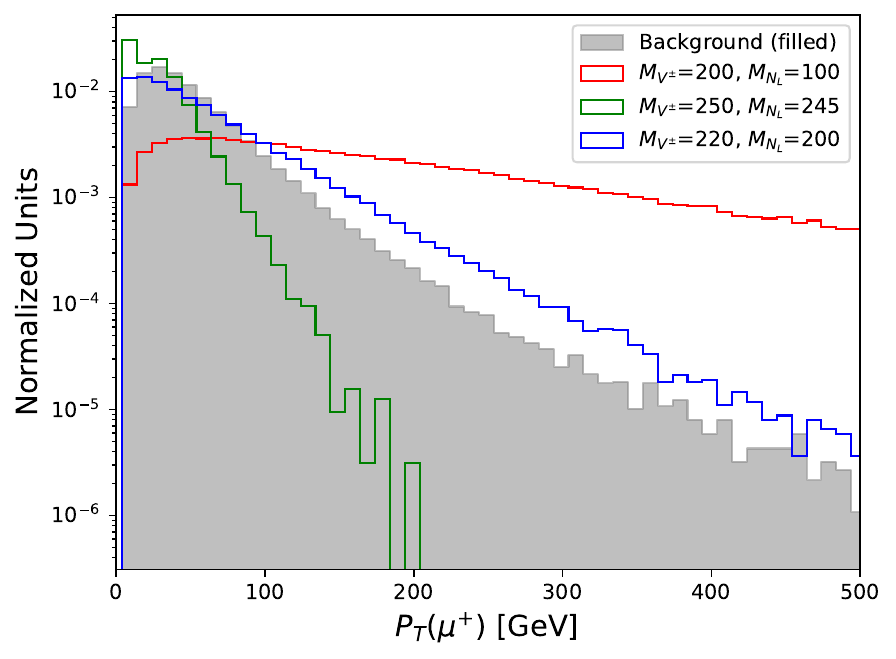}
        \label{fig:antimuon_pt_05}
    }
    
    \vspace{0.5cm} 
    \subfloat[]{
        \includegraphics[width=0.48\textwidth]{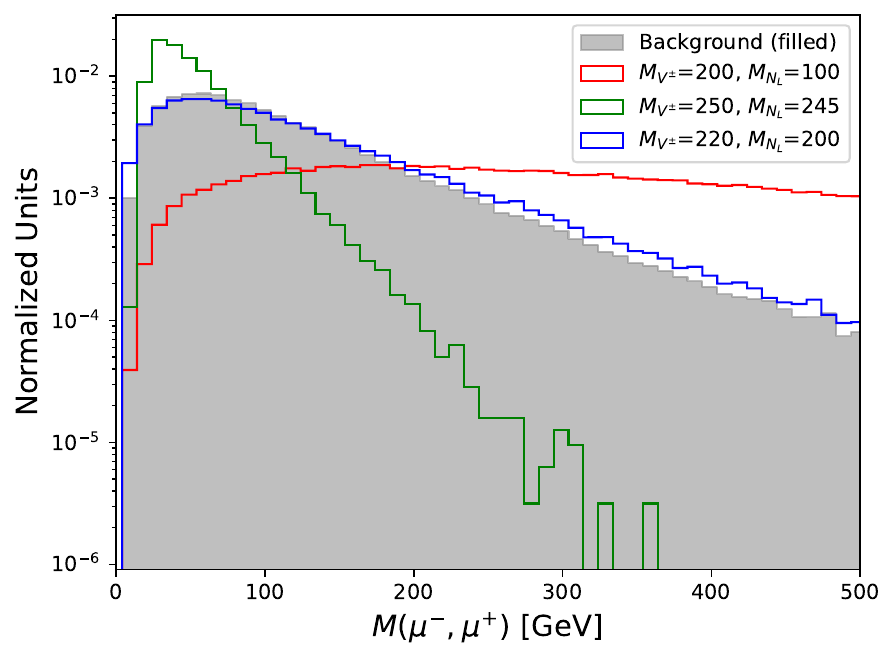}
        \label{fig:invmass_05}
    }
    \hfill
    \subfloat[]{
        \includegraphics[width=0.48\textwidth]{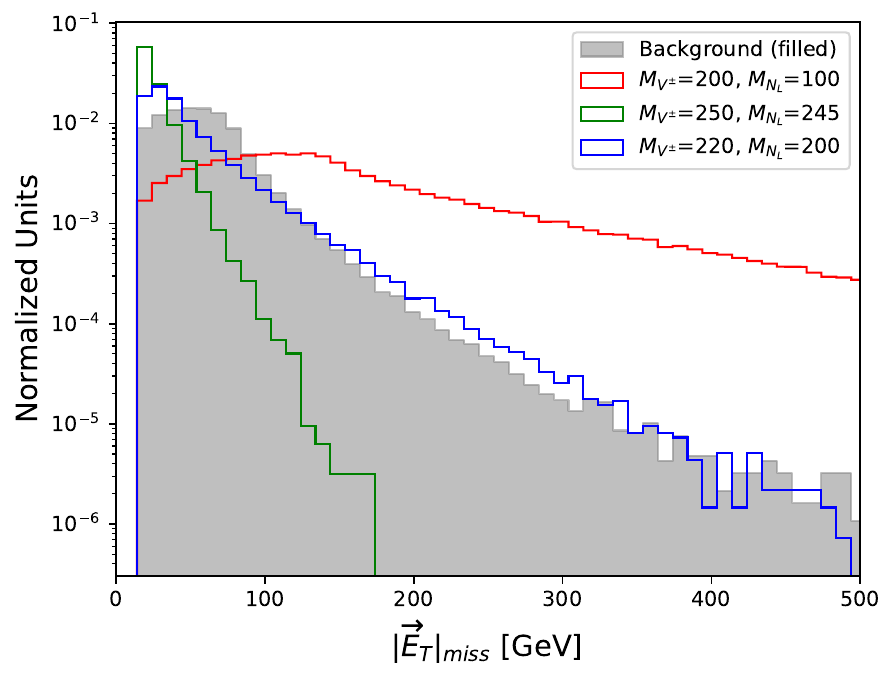}
        \label{fig:met_05}
    }
    \caption{Kinematic distributions for three mass scenarios and the background in pp-UPCs at the HL-LHC with $\sqrt{s}=14$ TeV,  where (a) is the muon transverse momentum, (b) is the anti-muon transverse momentum, (c) is the invariant mass of a pair of muons, and (d) is the missing transverse energy. The mass values are given in GeV.}
    \label{fig:kinematic}
\end{figure*}

Fig. \ref{fig:angular} show three different mass scenarios along with the background distributions for acoplanarity $\alpha$, separation in the $\eta \times \phi$ plane, and the cosine of the separation angle $\theta$ of muon pairs. Fig. \ref{fig:kinematic} presents the distributions for the muon/anti-muon transverse momentum $p_T$, the invariant mass $M$ of $\mu^{-}\mu^{+}$ pairs, and the missing transverse energy $|\overrightarrow{E}_T|_{\text{miss}}$.  \newline

Fig. \ref{fig:acoplanarity} shows significant differences between all mass scenarios and the background distribution, especially for values close to $\alpha \sim 0$. This implies that the muons tend to be produced back-to-back. The feature is particularly pronounced for the compressed-mass scenarios, providing a clear distinction from the background, which is broader and less pronounced. Similarly, the $\Delta R$ separation in Fig. \ref{fig:delta_r_separation_lhc_05} exhibits back-to-back kinematics for the muons, with a peak at $\Delta R \sim \pi$ for all mass scenarios shown; again, the compressed-mass scenarios are most favored. The $\cos\theta$ distributions (Fig. \ref{fig:costheta_mu1_mu2}) for all signal scenarios show a strong concentration in the negative region, sharply peaking near $\cos\theta \sim -1$, a feature that reflects the spin-1 nature of $V^{\pm}$. Regarding kinematic distributions, the behavior varies according to the mass hierarchy of $V^{\pm}$ and $N_{L}$. The transverse momentum distributions for both muons and antimuons (Fig. \ref{fig:muon_pt_05} and Fig. \ref{fig:antimuon_pt_05}, respectively) show that muons/anti-muons are significantly softer in compressed-mass scenarios, with their $p_T$ distributions peaking at much lower values and falling off rapidly, contrasting with the harder background.  For less compressed-mass scenarios, muons can be harder, with $p_T$ distributions extending to higher values than the background. This same behavior is shown in the invariant mass distribution (Fig. \ref{fig:invmass_05}) where compressed-mass scenarios show a prominent peak at very low invariant masses, a direct consequence of the soft muons produced in the $V^{\pm}$ decays and for less compressed-mass scenarios, the distribution extends beyond the background. The missing transverse energy distribution $|\overrightarrow{E}_T|_{\text{miss}}$ (Fig. \ref{fig:met_05}) is highly sensitive to $M_{N_L}$ and $M_{V^\pm}$. In compressed-mass scenarios, MET is very soft because HNLs carry little kinetic energy, unlike SM neutrinos, which makes the background distribution broader. For less compressed-mass scenarios with a lighter HNL, the MET distribution becomes harder and can extend to higher values than the background.   \newline

All these information are passed to the random search algorithm in order to find, for different ($M_{V^{\pm}},M_{N_L}$) mass scenarios in the 5 GeV$<M_{V^{\pm}},M_{N_{L}}<$600 GeV mass ranges, the best angular and kinematic cuts where statistical significance is maximum. The results of these cuts for some scenarios can be seen in Table \ref{table:table_atlas_cuts_hllhc}. This table shows that by sequentially applying the kinematic and angular cuts specified in each column, a higher proportion of signal events can be preserved relative to background events for the HL-LHC. This enables the identification of favorable regions in the parameter space for detecting new physics in potential HL-LHC searches.  \newline

\begin{table}[ht]
\begin{center}
\begin{tabular}{|c|c|c|c|c|c|c|c|c|c|c|c|} 
\hline
${M}_{{V}^{\pm}}$ & ${M}_{{N}_{L}}$ & $cos\theta_{\mu\mu}$ & $\alpha_{\mu\mu}<$ & $\Delta R_{\mu\mu}$ & $M_{\mu\mu} >$ & $|{\overrightarrow{E}}_{T}|_{miss} >$ & ${{p}_{T}}_{\mu}>$ & $N_{s}$ & $\epsilon^{cut}_{s}$ (\%) & $N_{b}$ & $\epsilon^{cut}_{b}$ (\%) \\
\hline
120 & 60 & -0.80$\pm$0.20 & 0.091 & 1.41$\pm$2.24 & 149.5 & 186.7 & 119.5 & 161 & 5.8 & 1 & 0.03 \\
\hline  
120 & 100 & -0.80$\pm$0.19 & 0.106 & 1.99$\pm$1.60 & 136.3 & 23.10 & 6.28 & 446 & 16 & 1 & 1.5 \\
\hline
140 & 40 & -0.80$\pm$0.19 & 0.101 & 0.91$\pm$2.68 & 150.2 & 197.8 & 121.1 & 105 & 7.5 & 1 & 0.03 \\
\hline
140 & 100 & -0.80$\pm$0.2 & 0.109 & 2.10$\pm$1.48 & 149.1 & 125.3 & 118.7 & 63 & 4.5 & 1 & 0.06 \\
\hline
160 & 80 & -0.80$\pm$0.19 & 0.118 & 2.71$\pm$0.83 & 150.4 & 173.3 & 123.6 & 52 & 6.9 & 1 & 0.03 \\
\hline
160 & 120 & -0.8$\pm$0.20 & 0.125 & 2.86$\pm$0.73 & 150.5 & 85.20 & 117.9 & 42 & 5.6 & 1 & 0.12 \\
\hline
180 & 80 & -0.81$\pm$0.20 & 0.140 & 1.74$\pm$1.84 & 150.1 & 195.4 & 119.4 & 31 & 7.2 & 1 & 0.03 \\
\hline
180 & 120 & -0.80$\pm$0.20 & 0.140 & 2.86$\pm$0.73 & 149.3 & 158.6 & 122.1 & 23 & 5.1 & 1 & 0.04 \\
\hline
200 & 100 & -0.80$\pm$0.19 & 0.156 & 1.28$\pm$2.28 & 149.2 & 188.6 & 119.6 & 19 & 7.33 & 1 & 0.03 \\
\hline
200 & 180 & -0.80$\pm$0.20 & 0.184 & 1.16$\pm$2.38 & 50.10 & 15.20 & 4.34 & 66 & 25 & 5 & 6.21 \\
\hline
\end{tabular}
\caption{Best angular and kinematic cuts for some (${M}_{{V}^{\pm}}$,${M}_{{N}_{L}}$) scenarios in the HL-LHC with pp-UPC. The transverse momentum, invariant mass, and missing transverse energy are given in GeV.}
\label{table:table_atlas_cuts_hllhc}
\end{center}
\end{table}

\begin{figure*}[h!]
    \centering
    \subfloat[]{
        \includegraphics[width=0.48\textwidth]{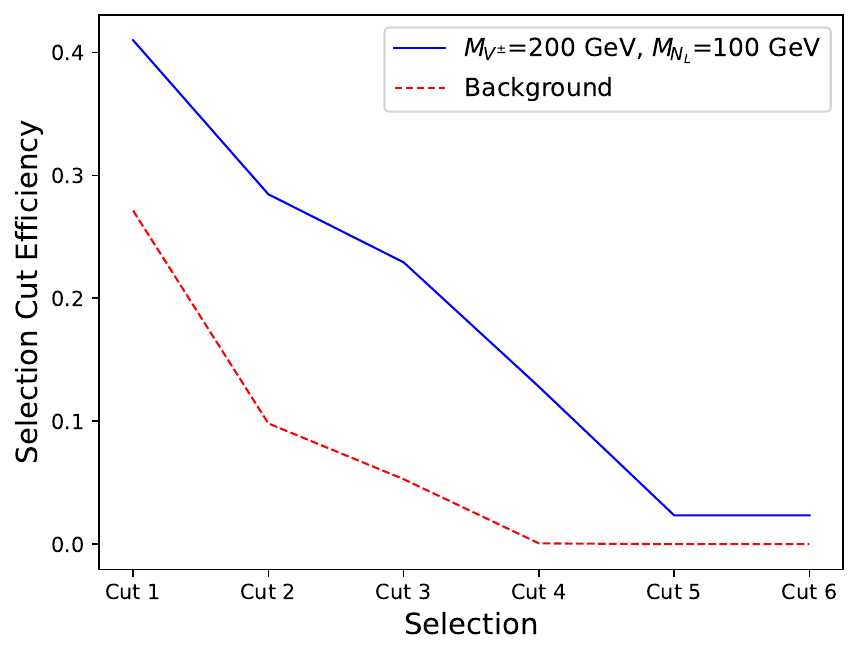}
        \label{fig:cutflow_signal1}
    }
    \hfill
    \subfloat[]{
        \includegraphics[width=0.48\textwidth]{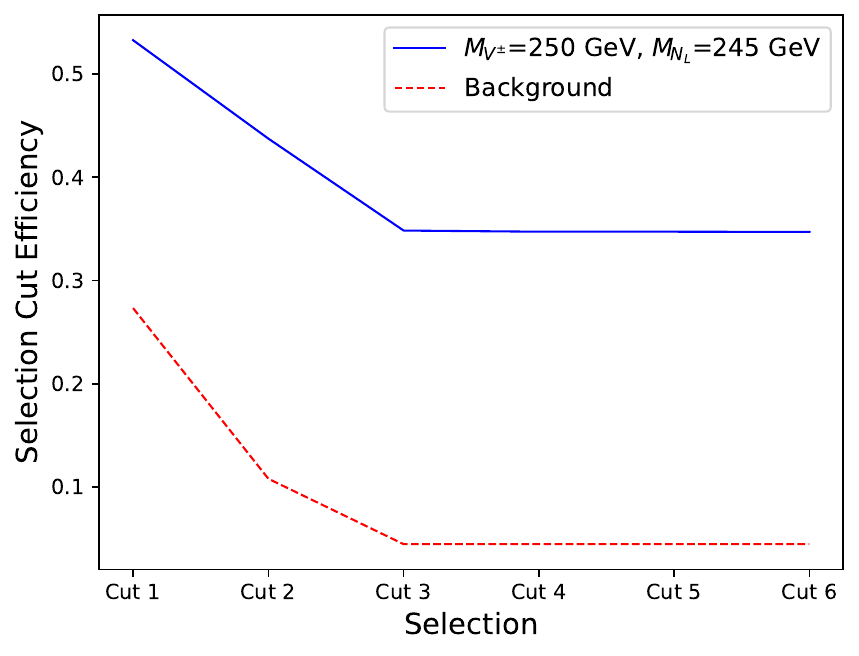}
        \label{fig:cutflow_signal5}
    }
    
    \vspace{0.5cm} 
    \subfloat[]{
        \includegraphics[width=0.48\textwidth]{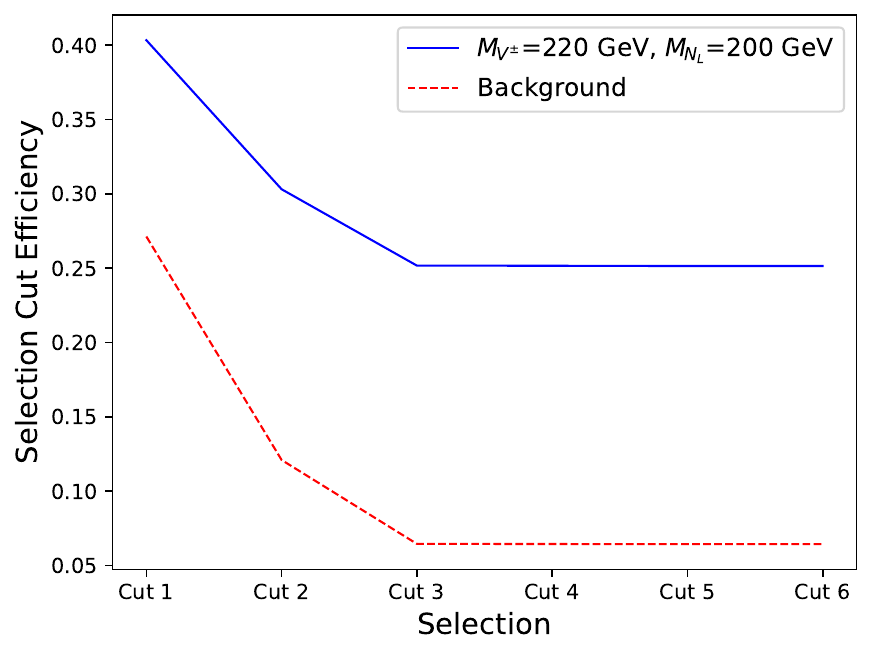}
        \label{fig:cutflow_signal10}
    }
    \caption{Cut Efficiency across each selection cut between the background and three different scenarios: (a) case when ${M}_{{V}^{\pm}} = 200$ GeV, ${M}_{{N}_{L}} = 100$ GeV; (b) case when ${M}_{{V}^{\pm}} = 250$ GeV, ${M}_{{N}_{L}} = 245$ GeV; (c) case when ${M}_{{V}^{\pm}} = 220$ GeV, ${M}_{{N}_{L}} = 200$ GeV. The selection cuts are show in Table \ref{table:table_atlas_cuts_hllhc}.}
    \label{fig:cutflow}
\end{figure*}

Complementary to Table \ref{table:table_atlas_cuts_hllhc}, Fig. \ref{fig:cutflow} shows the cut-flow efficiency for the background and three mass scenarios. In this plot we can see that the observables used and cuts found by the algorithm were optimal since they allowed a clear separation between signal and the background events.  \newline

\begin{figure*}[h!]
    \centering
    \subfloat[]{
        \includegraphics[width=0.48\textwidth]{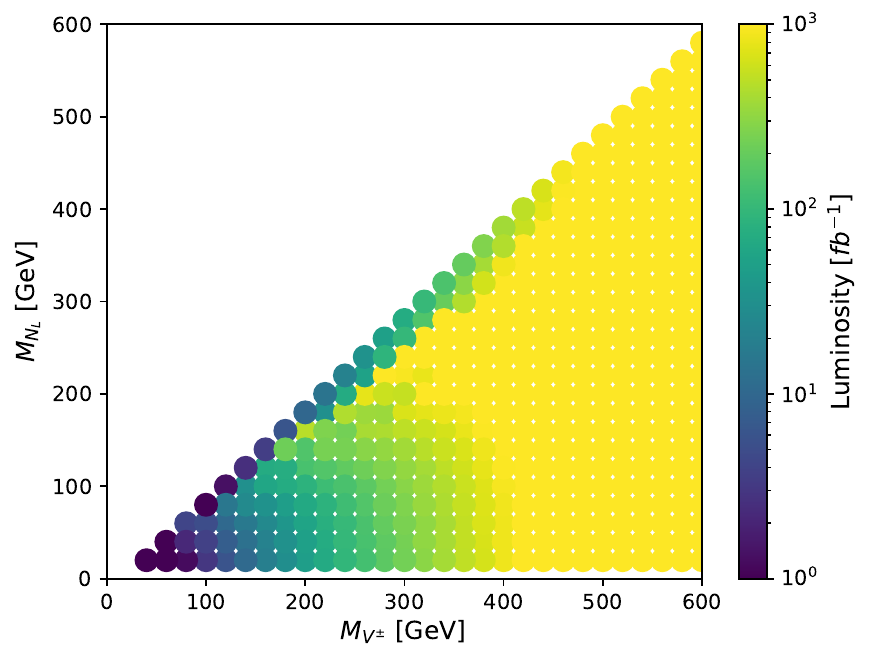}
        \label{fig:atlas_upc_lumi_2sigma}
    }
    \hfill
    \subfloat[]{
        \includegraphics[width=0.48\textwidth]{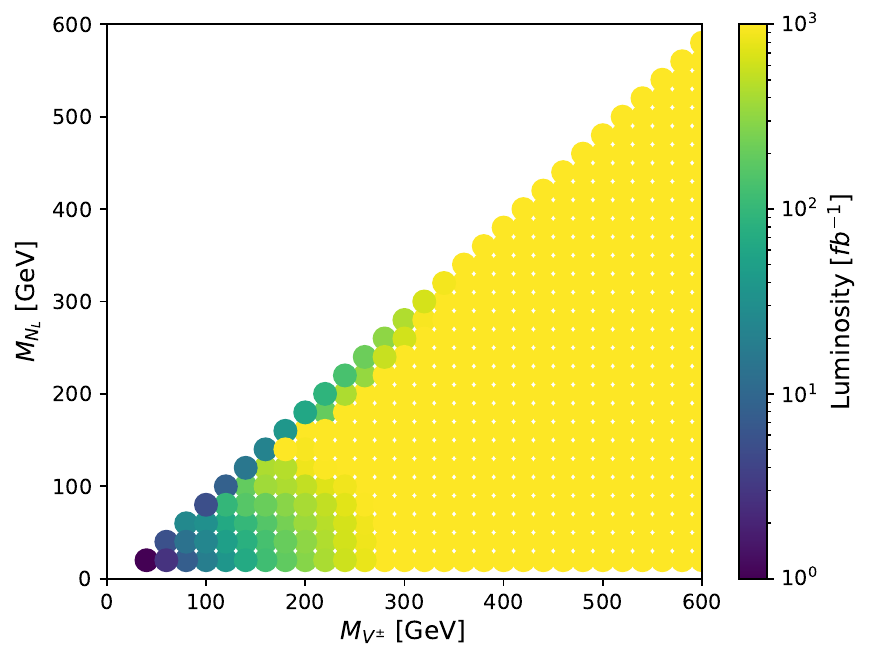}
        \label{fig:atlas_upc_lumi_5sigma}
    }
    
    \caption{Luminosity, in ${fb}^{-1}$, that the LHC at $\sqrt{s_{NN}}=14$ TeV must reach with UPCs of protons (a) to exclude a new charged vector boson at 95\% C.L. and (b) to discover with 5$\sigma$.}
    \label{fig:lumi_pp_upc}
\end{figure*}

The results of the optimized event selection analysis can be translated into obtaining sensitivity projections for the HL-LHC. Figure \ref{fig:lumi_pp_upc} shows the luminosity required in HL-LHC with pp-UPCs to exclude with 95\% CL or discover with 5$\sigma$ the different ($M_{V^{\pm}},M_{N_L}$) scenarios within the 5 GeV $<M_{V^{\pm}},M_{N_{L}}<$ 600 GeV mass range. We can observe that, with the expected luminosity for the HL-LHC with pp-UPCs, many scenarios in the mass range 5 GeV $<M_{V^{\pm}},M_{N_{L}}< 350$  GeV could be excluded with 95\% CL. Furthermore, in a discovery scenario, mass scenarios in the 5 GeV $<M_{V^{\pm}},M_{N_{L}}<$ 250 GeV mass range could reach 5$\sigma$ of statistical significance. In both cases, higher integrated luminosity is required to exclude or probe heavier mass scenarios, as expected.  \newline

\begin{figure*}[h!]
    \centering
    \subfloat[]{
        \includegraphics[width=0.48\textwidth]{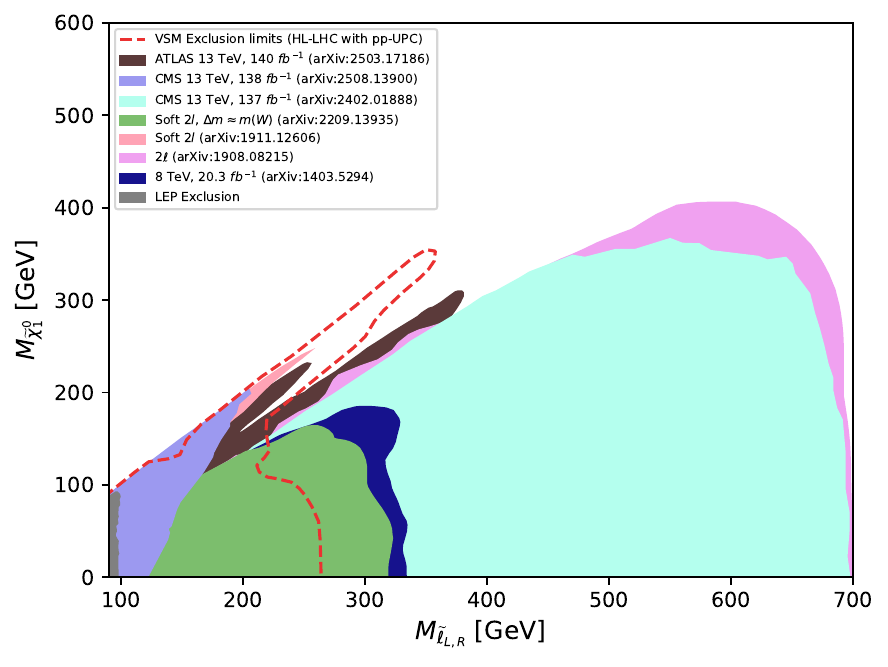}
        \label{fig:vsm_vs_MSSM1}
    }
    \hfill
    \subfloat[]{
        \includegraphics[width=0.48\textwidth]{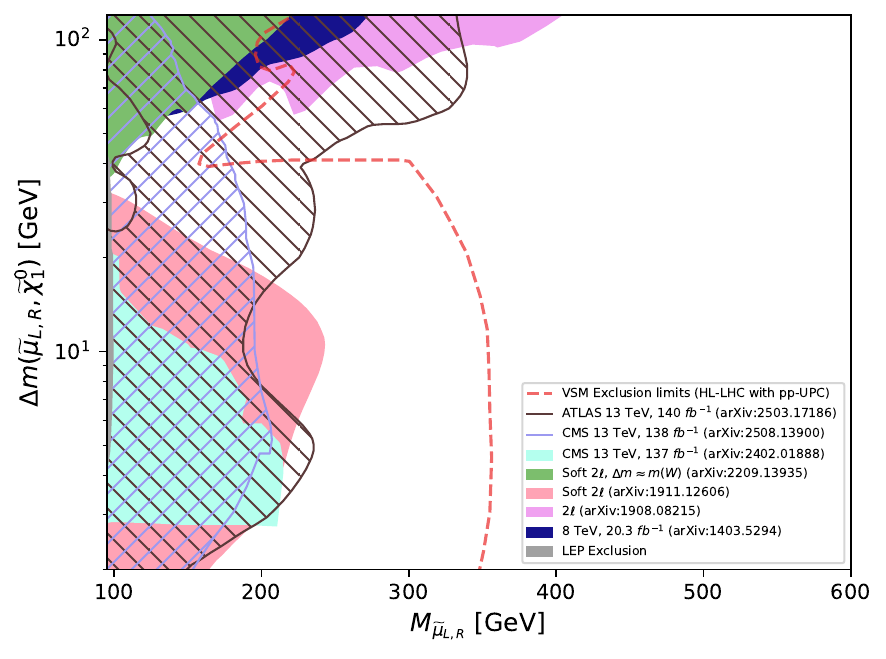}
        \label{fig:vsm_vs_MSSM2}
    }
    \caption{Projected 95\% C.L. exclusion limits for the VSM model using ultra-peripheral proton-proton collisions at the HL-LHC (red dashed line) against existing 95\% C.L. exclusion limits from SUSY searches at ATLAS, CMS, and LEP (colored regions) in the (a) $M_{\tilde{\ell}^{\pm}}$ $\times$ $M_{\tilde{\chi}_{1}^{0}}$ and (b) $M_{\tilde{\ell}^{\pm}}$ $\times$ $\Delta m(\tilde{\ell}^{\pm}, \tilde{\chi}_{1}^{0})$ mass plane.}
    \label{fig:vsm_vs_MSSM}
\end{figure*}

The dimuon+MET final states considered here are topologically similar to signatures predicted by many SUSY extensions. Specifically, the direct production of slepton pairs that decay into SM leptons and neutralinos, where the neutralinos escape the detector, produces the same observable final state. In particular, in our model, the vector boson $V^{\pm}$ and the HNL play a role similar to the smuon $\tilde{\mu}^{\pm}$ and the lightest neutralino $\tilde{\chi}_{1}^{0}$, respectively. In this context, for a first estimate, we will compare our results again with the exclusion limits for sleptons and neutralinos, now established by the ATLAS and CMS collaborations for the mass range analyzed here. However, it is crucial to note that while the slepton is a scalar particle, the vector boson $V^{\pm}$ studied in this work carries spin-1. This is very important because angular distributions of final-state particles is a powerful probe of the underlying physics, as it directly reflects the spin and coupling structure of the intermediate particles involved in the reaction. Consequently, different models should predict distinct angular distributions because they alter either the production mechanism or the spin correlations that govern the decay kinematics. These differences manifest as modifications in the acoplanarity $\alpha$, the separation in the $\eta \times \phi$ plane, and the cosine of the separation angle $\theta$  between the two muons, as illustrated in Figure \ref{fig:angular_lead}. Fore more details see paper ''Spin before mass at the LHC'' in reference \cite{Melia:2011cu}. In this context, Fig. \ref{fig:vsm_vs_MSSM} shows the projected exclusion limits of charged vector bosons and HNLs at the HL-LHC with pp-UPC, compared with the exclusion limits from SUSY searches of sleptons and neutralinos.  \newline

Figure \ref{fig:vsm_vs_MSSM1} presents the comparison between exclusion limits in the $M_{\tilde{\ell}^{\pm}}$ $\times$ $M_{\tilde{\chi}_{1}^{0}}$ mass plane, showing the reach of HL-LHC searches with pp-UPCs and existing SUSY limits from ATLAS, CMS, and LEP \cite{ALEPH:2002nwp, lepsusy2001, ATLAS:2025evx, ATLAS:2022hbt, ATLAS:2019lng, ATLAS:2019lff, ATLAS:2014zve}. The colored regions represent current experimental exclusion limits for direct slepton and neutralino production. Our projected exclusion limits (red dashed line) show a significant improvement in sensitivity: the HL-LHC with pp-UPC can probe regions of the parameter space where $M_{V^{\pm}}$ ranges from approximately 100 GeV up to around 400 GeV, and $M_{N_L}$ extends up to roughly 350 GeV. This again highlights the potential of UPC collisions to explore mass scenarios in the higher mass range, where previous SUSY searches established indirect constraints. Fig. \ref{fig:vsm_vs_MSSM2} complements these results by presenting the exclusion limits in the $M_{\tilde{\ell}^{\pm}}$ $\times$ $\Delta m(\tilde{\ell}^{\pm}, \tilde{\chi}_{1}^{0})$ mass plane. \newline 

Conventional proton-proton collisions in standard LHC searches often face significant challenges in compressed-mass regions due to the difficulties associated with triggering on soft leptons and low missing transverse energy. In contrast, our projected limits demonstrate strong sensitivity to these compressed-mass scenarios across a wide range of $M_{V^{\pm}}$ values. This capability stems from the significantly cleaner experimental environment provided by ultra-peripheral collisions, which effectively mitigate the hadronic backgrounds that typically obscure such soft signals. Consequently, the pp-UPC offers a unique window into these challenging regions, providing a powerful avenue for discovery or exclusions in regions where other searches have little sensitivity. It is important to re-emphasize that while this comparison is made to topologically similar SUSY processes, the distinct spin-1 nature of the $V^{\pm}$ boson in our model implies different angular distributions of the decay products. These angular signatures could be further exploited to enhance signal discrimination, potentially yielding even stronger sensitivity in direct searches of $V^{\pm}$ compared to slepton searches.  \newline

\section{Conclusions}
\label{conclusions}

In this work, we have investigated the potential of ultra-peripheral collisions at the LHC to probe new physics BSM, focusing on the production of new charged vector bosons that decay into heavy neutral leptons predicted by the Vector Scotogenic Model. In the first part of this study, we have demonstrated that the ATLAS experiment with ultra-peripheral Pb–Pb collisions at $\sqrt{s_{NN}}=5.02$ TeV and an integrated luminosity of $\mathcal{L}_{\text{int}}$ = 3.48 ${nb}^{-1}$ can directly search for a new charged vector boson in a mass region where LEP-II did not perform dedicated searches. We identified regions in the parameter space where the signal can be distinguished from the background with high statistical significance. In particular, we found that, PbPb-UPCs in ATLAS can exclude at 95\% C.L. mass scenarios such as $(M_{V^{\pm}},M_{N_L}) = (30,20)$ GeV, $(30,10)$ GeV and $(20,10)$ GeV, while a discovery significance of $5\sigma$ can be achieved for the $(20,10)$ GeV case. These results show that existing Pb–Pb data already provide direct sensitivity to light charged vector bosons below the indirect LEP-II limit inferred from low-mass slepton searches.  \newline

In the second part, we extended our analysis to ultra-peripheral proton–proton collisions at the HL-LHC, considering $\sqrt{s}=14$ TeV and $\mathcal{L}_{\text{int}}$ = 150 ${fb}^{-1}$. We have shown that the HL-LHC can exclude with 95\% C.L. a broad range of scenarios in the 100 GeV $< M_{V^{\pm}},M_{N_L} <$ 350 GeV mass range, and that most of this region can reach of $5\sigma$ of statistical significance. We have also compared our projections with existing bounds from slepton–neutralino searches at LEP, ATLAS and CMS. We find that pp-UPCs at the HL-LHC would cover regions of parameter space that extend beyond current direct slepton limits, in particular in compressed-mass regions that are challenging for conventional proton–proton analyses.  \newline

Our results show that ultraperipheral collisions offer a powerful and complementary way to explore physics beyond the Standard Model, providing a competitive sensitivity with respect to traditional searches in weakly constrained or completely unexplored regions. In particular, for the Scotogenic Vector Model, existing Pb-Pb data already allow direct tests of charged vector bosons in mass regions not explored by LEP, while future pp-UPC data at the HL-LHC can investigate masses up to a few hundred GeV with high sensitivity. 

\section*{Acknowledgements}

The work of Y.M. Oviedo-Torres and J. Zamora-Saa was funded by ANID - Millennium Science Initiative Program - ICN2019\_044. J. Zamora-Saa was partially supported by FONDECYT grant 1240216 and 1240066. Y.V. Oviedo-Torres was supported by FONDECYT grant 3250068.  The work of S. Tapia has been supported by ANID PIA/APOYO AFB230003.

\bibliography{referencias}

\end{document}

%% file: Diagrams.tex
\usepackage[scientific-notation=true]{siunitx}
\usepackage{float, verbatim, amsmath, amssymb, fullpage, paralist, cancel, listings, subfig, enumitem, siunitx, graphicx, xcolor, tikz, booktabs, tcolorbox, textcomp, url, parse lines, fontawesome5, marginnote, smartdiagram, forest, pifont, pgfcalendar, pgfgantt, fix-cm, ulem, soul, multirow, xfrac, colortbl, xspace, fontenc, helvet, mathptmx,ragged2e,csquotes} 
\definecolor{smcolor}{rgb}{0.8,0.5,0.5}
\definecolor{bsmcolor}{rgb}{0.4,0.6,0.8}
\definecolor{pvcolor}{rgb}{0.8,0.8,0.8}
\definecolor{cobalt}{rgb}{0.0, 0.28, 0.67}
\definecolor{darkelectricblue}{rgb}{0.33, 0.41, 0.47}
\definecolor{darkpowderblue}{rgb}{0.0, 0.2, 0.6}
\definecolor{darktangerine}{rgb}{1.0, 0.66, 0.07}
\definecolor{pastelviolet}{rgb}{0.8, 0.6, 0.79}
\definecolor{black}   {RGB}{0.0, 0.13, 0.28}
\definecolor{dukeblue}    {rgb}{0.0, 0.0, 0.61}
\definecolor{oxfordblue}{rgb}{0.0, 0.13, 0.28}
\definecolor{navyblue}{rgb}{0.0, 0.0, 0.5}
\definecolor{linkred}{RGB}{165,0,33}

\usesmartdiagramlibrary{additions}
\usepackage{hyperref}
\usepackage{cleveref}[capital]
\usepackage{textcase}
\hypersetup{
        linktocpage=true,
        setpagesize=true,
	colorlinks= true,
	linkcolor=cobalt!50!black,
        menucolor=cobalt!50!black,
	filecolor=cobalt!50!black,      
	urlcolor=cobalt!50!black,
        citecolor=cobalt!50!black,
        citebordercolor=cobalt!50!black,
	pdftitle={Light Dark Scalar},
        pdfsubject={Latex},
        pdfauthor={Yoxara Villamizar},
        pdfkeywords={BSM Phenomenology, SM, Collider, DM}
}
\graphicspath{{Figures/}} 
\newlist{firstitemize}{itemize}{1}
\setlist[firstitemize,1]{label=\color{cobalt!70!black} \ding{111}, nosep, leftmargin=*, before=\normalcolor}
\newlist{seconditemize}{itemize}{1}
\setlist[seconditemize,1]{label=\color{cobalt!70!black}\ding{226}\normalcolor, nosep, leftmargin=*} 
\newlist{thirditemize}{itemize}{1}
\setlist[thirditemize,1]{label=\color{cobalt!70!black} \ding{118}, nosep, leftmargin=*, before=\normalcolor}

\usepackage{tikz}
\usetikzlibrary{arrows.meta,decorations,calc,shadows.blur,decorations.pathreplacing,decorations.pathmorphing,decorations.markings,shapes, arrows, positioning,bending,chains,matrix, patterns,mindmap,trees}
\tikzset{
>=triangle 45,
scalar/.style={decorate, dashed, draw=cobalt!50!black, thick, line width=0.8pt}, 
photon/.style={decorate, line width=0.85pt, draw=black, decoration={snake, segment length=5, amplitude=3pt, post length=0.5mm, pre length=0.5mm}}, 
particle/.style={draw=black, postaction={decorate}},
fermion/.style={draw=black, line width=0.6pt, postaction={decorate}, decoration={markings,mark=at position .5 with {\arrow[draw=black,scale=1.2,>=stealth]{>}}}},
antifermion/.style={draw=black, line width=0.6pt, postaction={decorate},decoration={markings,mark=at position .5 with {\arrow[draw=black,scale=1.2,>=stealth]{<}}}},
}